%% file: OutflowsinCMZ_Jan18.tex
\documentclass[twocolumn,times]{aastex63}

\usepackage{CJKutf8}

\usepackage{amsmath}
\usepackage{amssymb}
\usepackage{amsfonts}
\usepackage{graphicx}
\usepackage{wasysym}
\usepackage{multirow}
\usepackage{mdframed}
\usepackage[T1]{fontenc}

\hypersetup{linkcolor=violet,citecolor=blue,filecolor=cyan,urlcolor=violet}

\newcommand{\meth}{\mbox{CH$_3$OH}}

\newcommand{\tco}{\mbox{$^{13}$CO}}
\newcommand{\ceo}{\mbox{C$^{18}$O}}
\newcommand{\fmh}{\mbox{H$_2$CO}}
\newcommand{\httco}{\mbox{H$_2$$^{13}$CO}}
\newcommand{\mthc}{\mbox{CH$_3$CN}}
\newcommand{\cyacet}{\mbox{HC$_3$N}}

\newcommand{\water}{\mbox{H$_2$O}}

\newcommand{\kms}{\mbox{km\,s$^{-1}$}}

\newcommand{\sqc}{\mbox{cm$^{-2}$}}
\newcommand{\cc}{\mbox{cm$^{-3}$}}

\newcommand{\msol}{\mbox{$M_\odot$}}
\newcommand{\msolpyr}{\mbox{$M_\odot$\,yr$^{-1}$}}
\newcommand{\mujypbm}{\mbox{$\mu$Jy\,beam$^{-1}$}}
\newcommand{\mjypbm}{\mbox{mJy\,beam$^{-1}$}}
\newcommand{\jypbm}{\mbox{Jy\,beam$^{-1}$}}

\newcommand{\hii}{\mbox{H\,{\sc ii}}}
\newcommand{\vlsr}{\mbox{$V_\text{lsr}$}}

\newcommand{\ctw}{the 20~\kms{} cloud}
\newcommand{\cfi}{the 50~\kms{} cloud}
\newcommand{\Ctw}{The 20~\kms{} cloud}
\newcommand{\Cfi}{The 50~\kms{} cloud}

\newcommand{\sgb}{Sgr~B1-off}

\defcitealias{lu2020a}{Paper I}

\received{- -, --}
\revised{- -, --}
\accepted{- -, --}
\submitjournal{ApJ}

\shorttitle{Protostellar Outflows in CMZ Clouds}
\shortauthors{Lu et al.}

\begin{document}
\begin{CJK}{UTF8}{gbsn}

\title{ALMA Observations of Massive Clouds in the Central Molecular Zone: Ubiquitous Protostellar Outflows}
\correspondingauthor{Xing Lu}
\email{xinglv.nju@gmail.com, xing.lu@nao.ac.jp}

\author[0000-0003-2619-9305]{Xing Lu (吕行)}
\affiliation{National Astronomical Observatory of Japan, 2-21-1 Osawa, Mitaka, Tokyo 181-8588, Japan}

\author[0000-0003-1275-5251]{Shanghuo Li}
\affiliation{Korea Astronomy and Space Science Institute, 776 Daedeokdae-ro, Yuseong-gu, Daejeon 34055, Republic of Korea}
\affiliation{Shanghai Astronomical Observatory, Chinese Academy of Sciences, 80 Nandan Road, Shanghai 200030, P. R. China}
\affiliation{University of Chinese Academy of Sciences, 19A Yuquanlu, Beijing 100049, P. R. China}

\author{Adam Ginsburg}
\affiliation{Department of Astronomy, University of Florida, P.O.\ Box 112055, Gainesville, FL 32611, USA}

\author{Steven N.\ Longmore}
\affiliation{Astrophysics Research Institute, Liverpool John Moores University, IC2, 146 Brownlow Hill, Liverpool, L3 5RF, United Kingdom}

\author[0000-0002-8804-0212]{J.~M.~Diederik~Kruijssen}
\affiliation{Astronomisches Rechen-Institut, Zentrum f\" ur Astronomie der Universit\" at Heidelberg, M\"onchhofstra\ss e 12-14, D-69120 Heidelberg, Germany}

\author{Daniel L. Walker}
\affiliation{Department of Physics, University of Connecticut, 196A Auditorium Road, Storrs, CT 06269, USA}

\author{Siyi Feng}
\affiliation{National Astronomical Observatories, Chinese Academy of Science, Beijing 100101, P. R. China}
\affiliation{Academia Sinica Institute of Astronomy and Astrophysics, No.\ 1, Section 4, Roosevelt Road, Taipei 10617, Taiwan}
\affiliation{National Astronomical Observatory of Japan, 2-21-1 Osawa, Mitaka, Tokyo, 181-8588, Japan}

\author{Qizhou Zhang}
\affiliation{Center for Astrophysics | Harvard \& Smithsonian, 60 Garden Street, Cambridge, MA 02138, USA}

\author[0000-0002-6073-9320]{Cara Battersby}
\affiliation{Department of Physics, University of Connecticut, 196A Auditorium Road, Storrs, CT 06269, USA}

\author{Thushara Pillai}
\affiliation{Institute for Astrophysical Research, 725 Commonwealth Ave, Boston University Boston, MA 02215, USA}

\author{Elisabeth A.\ C.\ Mills}
\affiliation{Department of Physics and Astronomy, University of Kansas, 1251 Wescoe Hall Dr., Lawrence, KS 66045, USA}

\author{Jens Kauffmann}
\affiliation{Massachusetts Institute of Technology, 99 Millstone Road, Haystack Observatory, Westford, MA 01886, USA}

\author{Yu Cheng}
\affiliation{Department of Astronomy, University of Virginia, Charlottesville, VA 22904, USA}

\author{Shu-ichiro Inutsuka}
\affiliation{Department of Physics, Graduate School of Science, Nagoya University, Nagoya 464-8602 , Japan}

\begin{abstract} 
We observe 1.3~mm spectral lines at 2000~AU resolution toward four massive molecular clouds in the Central Molecular Zone of the Galaxy to investigate their star formation activities. We focus on several potential shock tracers that are usually abundant in protostellar outflows, including SiO, SO, \meth{}, \fmh{}, \cyacet{}, and HNCO. We identify 43 protostellar outflows, including 37 highly likely ones and 6 candidates. The outflows are found toward both known high-mass star forming cores and less massive, seemingly quiescent cores, while 791 out of the 834 cores identified based on the continuum do not have detected outflows. The outflow masses range from less than 1~\msol{} to a few tens of \msol{}, with typical uncertainties of a factor of 70. We do not find evidence of disagreement between relative molecular abundances in these outflows and in nearby analogs such as the well-studied L1157 and NGC7538S outflows. The results suggest that i) protostellar accretion disks driving outflows ubiquitously exist in the CMZ environment, ii) the large fraction of candidate starless cores is expected if these clouds are at very early evolutionary phases, with a caveat on the potential incompleteness of the outflows, iii) high-mass and low-mass star formation is ongoing simultaneously in these clouds, and iv) current data do not show evidence of difference between the shock chemistry in the outflows that determines the molecular abundances in the CMZ environment and in nearby clouds.
\end{abstract}


\section{INTRODUCTION}\label{sec:intro}
Star formation in the Central Molecular Zone (CMZ; the inner 500~pc of our Galaxy) has been a controversial topic. With more than $10^7$~\msol{} gas of mean density at $10^4$~\cc{} \citep{morris1996,ferriere2007,longmore2013a}, the CMZ exhibits about 10 times less efficient star formation than expected by dense gas-star formation relations that have been tested toward nearby molecular clouds and external galaxies \citep{longmore2013a,kruijssen2014,barnes2017}. Massive clouds in the CMZ have been suggested to be progenitors of young massive star clusters \citep{longmore2013b,rathborne2015,walker2016}, but observations reveal inefficient star formation in these clouds \citep{kauffmann2017a,walker2018,lu2019a,lu2019b}, with an overall dearth of compact dense cores across much of the CMZ \citep{cmzoom2020,hatchfield2020}.

In \citet[][hereafter \citetalias{lu2020a}]{lu2020a}, we reported ALMA Band~6 continuum observations toward four clouds, including \ctw{}, \cfi{}, \sgb{}, and Sgr~C, which are some of the most massive clouds in the CMZ and show signs of embedded star formation \citep{kauffmann2017a}. We identified hundreds of 2000 AU-scale cores in three of the clouds (the exception being \cfi{}), and suggested that the three clouds will likely form OB associations that contain less than 20 high-mass stars and have a spatial extent of $\sim$5--10~pc. In \cfi{}, no cores above the 5$\sigma$ level and larger than the synthesized beam were found, likely because this cloud has evolved to a much later phase when cold cores vanish and \hii{} regions dominate \citep{mills2011}. At sub-0.1~pc scales, we found evidence of thermal Jeans fragmentation and a similar core mass function as in Galactic disk clouds, which may hint at similar star formation processes at small spatial scales taking place in the CMZ and elsewhere in the Galaxy.

\begin{figure*}[!t]
\centering
\includegraphics[width=0.99\textwidth]{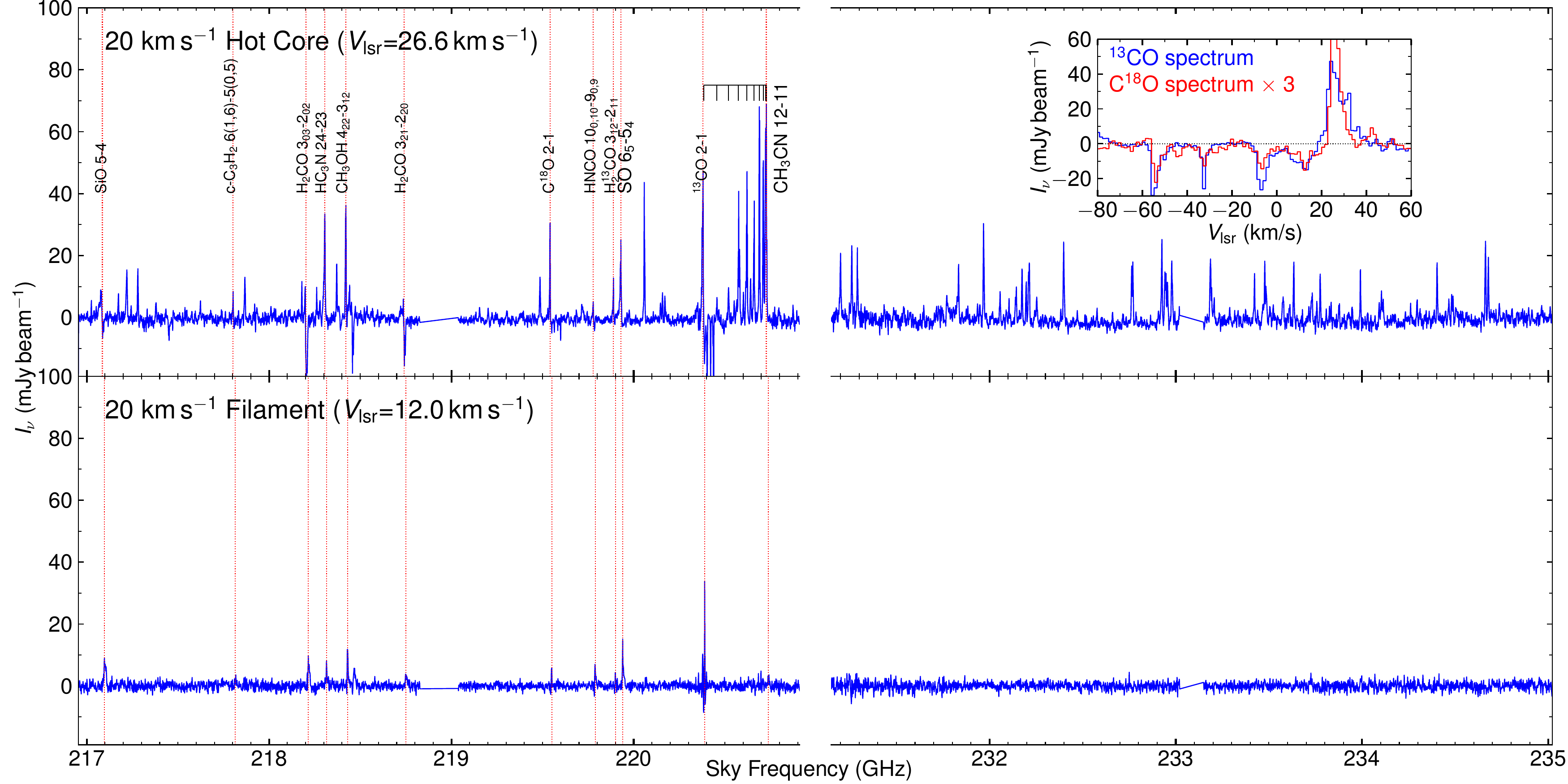} \\
\caption{Typical 1.3~mm spectra captured by ALMA, toward a star forming hot core and a chemically quiescent region spatially offset from star forming regions in \ctw{}, respectively. The corresponding positions where the spectra are extracted are denoted by blue arrows and labeled in \autoref{fig:c20kms}. The lines plotted in \autoref{fig:c20kms} are highlighted by vertical dashed lines. Most of the other lines detected toward the hot core are from rotational transitions of complex organic molecules. The inset shows the \tco{} and \ceo{} spectra toward the hot core along the \vlsr{} axis. Absorption at $-$55, $-$30, and $-$5~\kms{} owing to foreground gas is seen.}
\label{fig:spec}
\end{figure*}

However, it is unclear whether these cores are prestellar or protostellar (i.e., whether there are already embedded protostars). In \citet{lu2019a,lu2019b}, we used \water{} masers, class II \meth{} masers, and ultra-compact \hii{} regions to trace high-mass star formation and identified a few high-mass star forming regions in these clouds. Yet, the relatively poor resolution of those observations, $\sim$4\arcsec{} ($\sim$33,000 AU), prevents us from associating a particular high-mass star formation tracer with any of the 2000 AU-scale cores in \citetalias{lu2020a}. Further, the ultra-compact \hii{} regions trace a later evolutionary phase of high-mass star formation, such that we may miss low to intermediate-mass star formation or early evolutionary phases of high-mass star formation. The masers are able to reveal low to intermediate-mass star formation, but suffer from potentially low detection rates and contamination from masing sources other than star forming regions.

To this end, molecular outflows associated with the 2000 AU-scale cores are a promising tracer of star formation. Outflows are ubiquitously found in star forming regions, and are detected around both low-mass and high-mass protostars across a wide range of evolutionary phases as long as gas accretion is underway \citep[e.g.,][]{zhang2005,arce2010,li2019,lish2020}. Several molecular lines that are potential outflow tracers were observed along with the continuum data in \citetalias{lu2020a}. Therefore, in this paper we use the lines to search for direct evidence of star formation in the form of protostellar outflows.

In addition, the shock chemistry in protostellar outflows in the CMZ is poorly constrained. For one thing, only a handful of protostellar outflows have been detected in the CMZ, which are mostly in the most actively star forming region, Sgr~B2 \citep{qin2008,higuchi2015}. More recently, with the advent of high resolution ALMA observations, more outflows are being detected outside of Sgr~B2 (e.g., D.\ Walker et al.\ submitted, 2020). On the other hand, the chemistry of the molecular gas at pc scales in the CMZ seems to be distinct from that in the solar neighborhood, with noticeable enhancement of complex organic molecules and shock tracers throughout the CMZ \citep{martinpintado1997,requenatorres2006,requenatorres2008,menten2009}, likely caused by the extreme conditions such as widespread shocks, high gas temperatures, high cosmic ray fluxes, and high X-ray fluxes \citep{mills2013,ginsburg2016,henshaw2016,padovani2020,bykov2020}. Once we obtain a large sample of outflows, we will be able to systematically compare the relative abundances of the molecules in these outflows to those in nearby clouds, to investigate whether the shock chemistry differs between the CMZ and elsewhere in the Galaxy at sub-0.1 pc scales.

In the following, we first introduce our ALMA observation and data reduction strategies, as well as an assessment of the missing flux issue in \autoref{sec:obs}. Then in \autoref{sec:results} we summarize our observational results, including an overview of the line emission and a visual identification of outflows. In \autoref{sec:disc}, we estimate column densities, molecular abundances, and masses of the outflows, and discuss the implications to chemistry and star formation. We conclude our paper in \autoref{sec:conclusions}. In \autoref{appd_sec:column}, we introduce the procedures to estimate column densities using the molecular line data. In \autoref{appd_sec:outflowtable}, we list properties of the identified outflows in a detailed table. Throughout the paper we adopt a distance of 8.178~kpc to the CMZ \citep{gravity2018}.

\section{OBSERVATIONS AND DATA REDUCTION}\label{sec:obs}

\subsection{ALMA Observations}\label{subsec:obs_alma}
The ALMA observations were carried out in the C40-3 and C40-5 configurations in April and July of 2017 (project code: 2016.1.00243.S). Details of the sample selection, observation setup, and data calibration can be found in \citetalias{lu2020a}. The four clouds in the sample are listed in \autoref{tab:clouds}. The covered fields are chosen based on Submillimeter Array (SMA) and VLA observations that revealed potential sites of high-mass star formation including \water{} masers and massive dense cores of 0.2~pc scale \citep{lu2019a}. We imaged the Band 6 (1.3~mm) spectral lines using CASA 5.4.0. The covered frequencies range from 217--221~GHz to 231--235~GHz with a uniform channel width of 0.977~MHz (1.3~\kms{}). The effective spectral resolution is 1.129~MHz (1.5~\kms{}) after a Hanning smoothing done by the observatory.

We first manually identified line-free channels in the visibility data, and fed them to the \textit{uvcontsub} task to subtract the continuum baseline. Then we used the \textit{tclean} task to image the spectral lines, with Briggs weighting and a robust parameter of 0.5, and multi-scale algorithm with scales of [0, 5, 15, 50, 150] pixels and a pixel size of 0\farcs{04}. The image reconstruction was carried out in a two-step manner: first, the auto-masking algorithm with the recommended parameters\footnote{\url{https://casaguides.nrao.edu/index.php/Automasking_Guide}} was employed in \textit{tclean} to automatically identify and clean signals; then the \textit{tclean} task was restarted using clean models and residuals from the previous step as input, and all pixels within 20\% primary beam response included in the clean mask, to a threshold of $\sim$5$\sigma$ (8~\mjypbm{} per channel), in order to clean any residual significant signals. In a few cases where strong spatially diffuse emission is detected (e.g., \fmh{} in \ctw{}), a threshold of 8~\mjypbm{} may be too low and causes the clean algorithm to diverge. We elevated the threshold to 10~\mjypbm{} for these lines.  We have compared images produced by our automatic approach with those produced by a fine-tuned manual \textit{tclean} of several lines, and found that the images are almost identical.

The resulting synthesized beam is on average 0\farcs{28}$\times$0\farcs{19} (equivalent to 2200~AU$\times$1500~AU) but slightly varies between lines. The largest recoverable angular scale is 10\arcsec{} ($\sim$0.4~pc) with a shortest baseline length of 17~m. The achieved spectral line rms is between 1.6--2.0~\mjypbm{} (0.8--1.0~K in brightness temperatures) per 1.3~\kms{} channel depending on frequencies and regions.
 
\begin{deluxetable*}{ccccccccccc}
\tabletypesize{\scriptsize}
\tablecaption{Properties of the clouds.\label{tab:clouds}}
\tablewidth{0pt}
\tablehead{
\colhead{Cloud} & \vlsr{} & No.\ of cores & No.\ of outflows & Fraction with outflows & $\bar{X}$(SiO) & $\bar{X}$(SO) & $\bar{X}$(\meth) & $\bar{X}$(\fmh) & $\bar{X}$(\cyacet) & $\bar{X}$(HNCO) \\
 (1) & (2) & (3) & (4) & (5) & (6) & (7) & (8) & (9) & (10) & (11)
 }
\startdata
\Ctw{} & 12.5        & 471 & 20 & 0.042 & 1.69$\times$10$^{-9}$ & 7.10$\times$10$^{-9}$ & 0.99$\times$10$^{-7}$ & 0.80$\times$10$^{-8}$ & 0.65$\times$10$^{-9}$ & 4.40$\times$10$^{-9}$\\
\sgb{}  & 31.1        &  89  & 5   & 0.056 & 1.39$\times$10$^{-9}$ & 6.19$\times$10$^{-9}$ & 0.86$\times$10$^{-7}$ & 1.09$\times$10$^{-8}$ & 1.24$\times$10$^{-9}$ & 4.32$\times$10$^{-9}$\\
Sgr~C  & $-$52.6 & 274 & 18 & 0.066 & 2.39$\times$10$^{-9}$ & 9.87$\times$10$^{-9}$ & 1.33$\times$10$^{-7}$ & 1.84$\times$10$^{-8}$ & 1.48$\times$10$^{-9}$ & 5.44$\times$10$^{-9}$\\
\hline
All three clouds & \nodata & 834 & 43 & 0.052 & 2.05$\times$10$^{-9}$ & 8.49$\times$10$^{-9}$ & 1.15$\times$10$^{-7}$ & 1.40$\times$10$^{-8}$ & 1.16$\times$10$^{-9}$ & 4.95$\times$10$^{-9}$ \\
\hline
\Cfi{}   & 48.6       & 0 & 0 & \nodata & \nodata & \nodata & \nodata & \nodata & \nodata & \nodata  \\
\enddata
\tablecomments{Column (1): cloud name. Column (2): \vlsr{} of the cloud \citep{kauffmann2017a}. Column (3): number of the identified cores (\citetalias{lu2020a}). Column (4): number of the identified outflows in this work. Column (5): fraction of the cores that have identified outflows. Columns (6--11): mean molecular abundances of all the outflows in the cloud. Only the abundances with independent measurements are considered (entries without notes in the last column in \autoref{tab:outflows}; see \autoref{subsubsec:disc_abundance}).}
\end{deluxetable*}

\subsection{Assessing the Missing Flux}\label{subsec:obs_combine}
By their nature, interferometric observations do not recover structures on size scales larger than their largest recoverable angular scale (10\arcsec{} or 0.4~pc for these observations). If structures larger than this exist in the field, the flux captured by interferometers will be smaller than the true flux, which is referred to as the missing flux problem. Spatially extended ($\gtrsim$0.4~pc) structures including outflows and filaments are seen in the spectral line images. In principle, we can combine our data with shorter baseline as well as single-dish data to recover any spatially extended emission. Such data are available from the SMA and the Caltech Submillimeter Observatory (CSO) observations published in \citet{lu2017,lu2019a} and \citet{cmzoom2020}. 

However, we note several issues that prevent us from efficiently combining the data: i) The sensitivity of the SMA data is not optimal for the combination. For several regions, e.g., Sgr~C, the SMA observation recovers an even smaller flux than the ALMA data, suggesting that some weaker emission is missed by the SMA data due to its lower sensitivity. ii) Several regions of particular interest (e.g., the western part of Sgr~C) are not well sampled by existing SMA observations.

We attempted to combine the ALMA and SMA data, by concatenating the visibility data and imaging them together. The resulting image is not improved compared to the ALMA image, e.g., the rms becomes higher, and the integrated fluxes of several spatially extended structures do not increase significantly (in the extreme case of Sgr~C, the fluxes even decrease).

Therefore, we conclude that the imaging of diffuse structures does not benefit from the addition of the SMA data. This does not rule out the possibility that better shorter baseline data may help. A longer SMA observation, or an ACA observation that provides better sensitivity than the SMA, would be necessary to be combined with the ALMA data to recover spatially extended emission.

Meanwhile, we compared the integrated fluxes recovered by the CSO and ALMA, focusing on the SiO emission in Sgr~C which is the most spatially extended in our data and thus the most affected by the missing flux. In a circle of 50\arcsec{} diameter centered in between the two clusters in Sgr~C, the ALMA/CSO SiO integrated fluxes are measured to be 120/200 Jy\,\kms{}. The ALMA data recover 60\% of the flux observed by the CSO. We thus estimate an upper limit of 40\% for the missed flux in our ALMA data. In \autoref{subsubsec:disc_error}, we will see that this does not affect our estimate of outflow properties, as the dominant uncertainty is the molecular abundance that is unconstrained by several orders of magnitude.

\begin{figure*}[!t]
\centering
\includegraphics[width=0.95\textwidth]{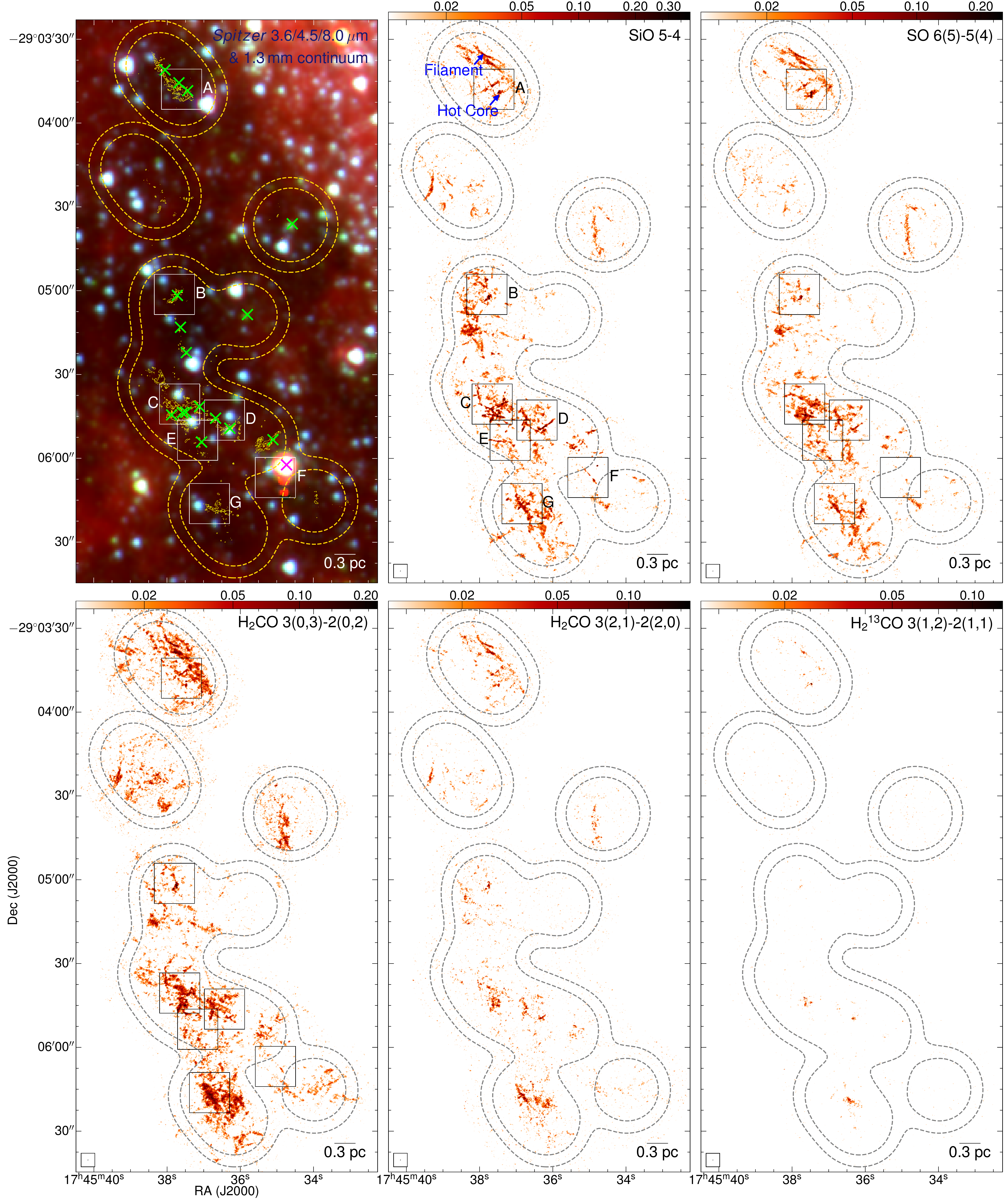}
\caption{Molecular line emission in \ctw{}. The inner and outer dashed loops in all panels demonstrate the 50\% and 30\% primary beam responses of the ALMA mosaics. The first panel shows a three-color image made from \textit{Spitzer} IRAC 3.6, 4.5, and 5.8~\micron{} bands, with yellow contours overlaid illustrating the ALMA 1.3 mm continuum emission at levels of [5,25,45]$\times$40~\mujypbm{}. Positions of \water{} masers are marked by crosses, among which those with AGB star counterparts \citep{lu2019a} are colored in magenta. The other panels show integrated intensities of molecular lines in a logarithmic scale, which are integrated in a velocity range of [$-20$,$40$] \kms{}, except for \meth{} and \mthc{} where this range is adjusted to avoid confusion with adjacent lines. The colorbars are in unit of \jypbm{}\,\kms{}. In selected panels, black boxes show regions where outflows are identified, with zoomed-in views presented in Figures~\ref{fig:outflows_c20_a}--\ref{fig:outflows_sgrc_g}.}
\label{fig:c20kms}
\end{figure*}

\addtocounter{figure}{-1}
\begin{figure*}[!p]
\centering
\includegraphics[width=0.95\textwidth]{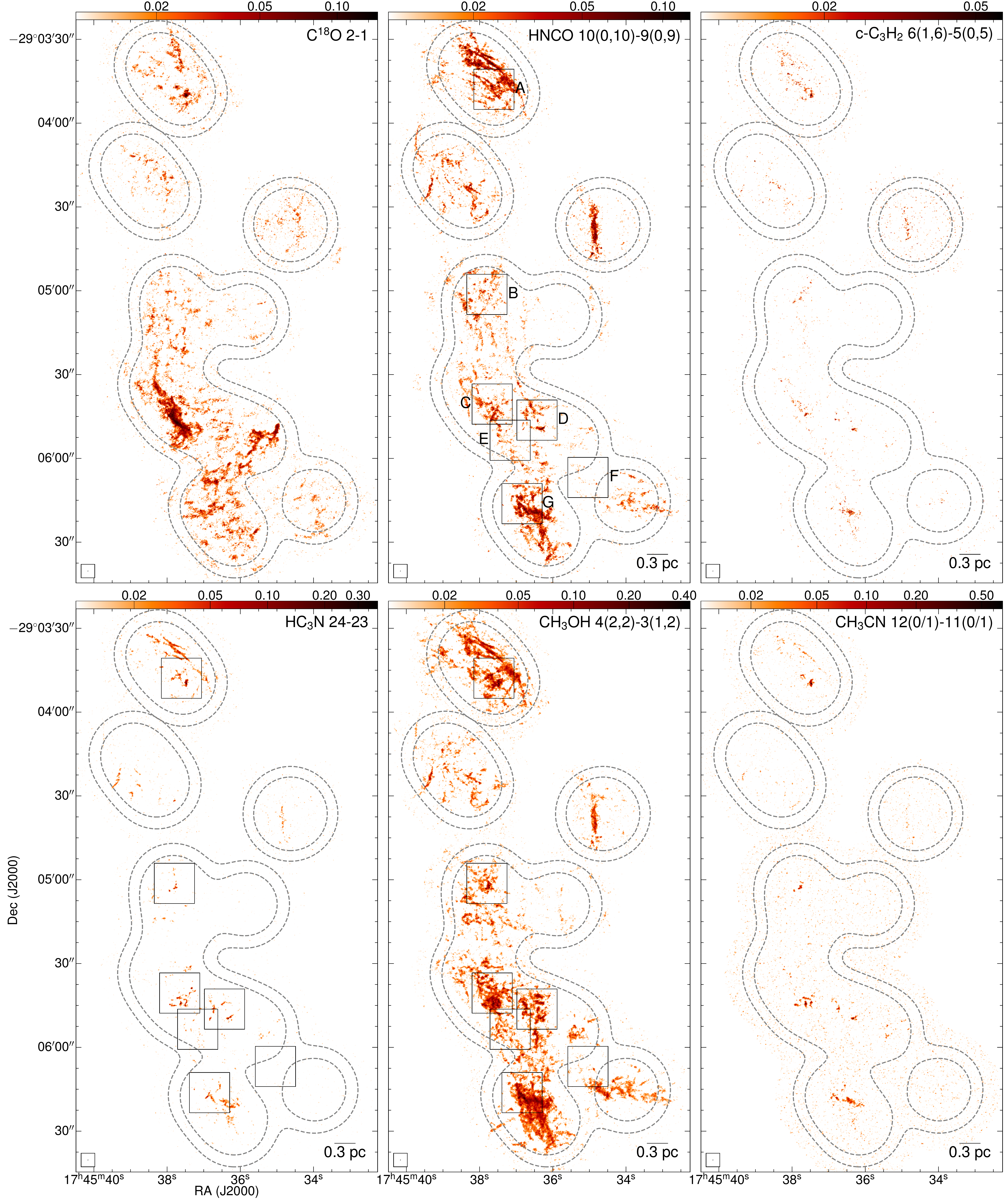}
\caption{Continued.}
\end{figure*}

\begin{figure*}[!p]
\centering
\includegraphics[width=0.95\textwidth]{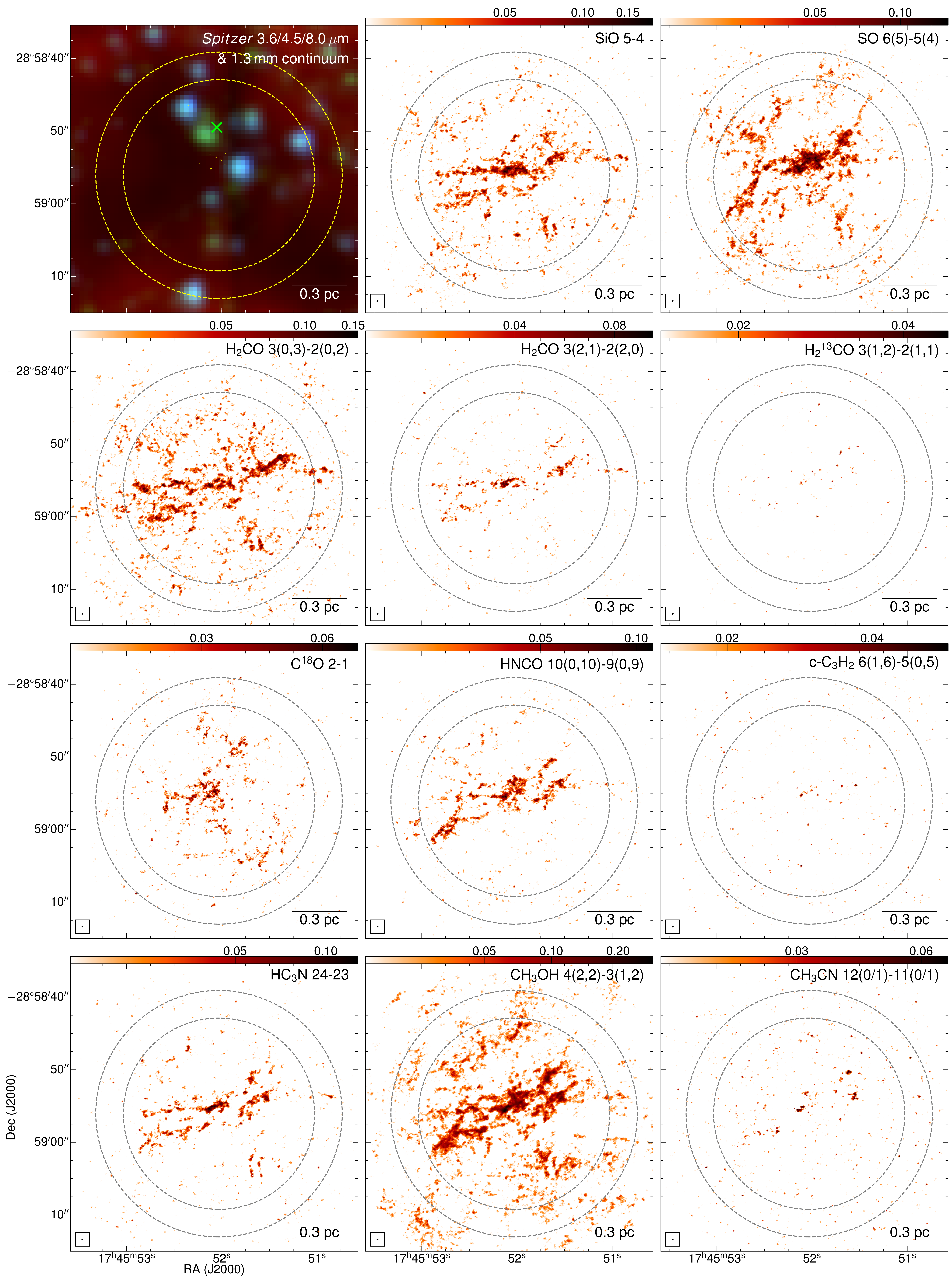}
\caption{Molecular line emission in \cfi{}. The layout of panels and symbols is the same as in \autoref{fig:c20kms}.}
\label{fig:c50kms}
\end{figure*}

\begin{figure*}[!p]
\centering
\includegraphics[width=0.95\textwidth]{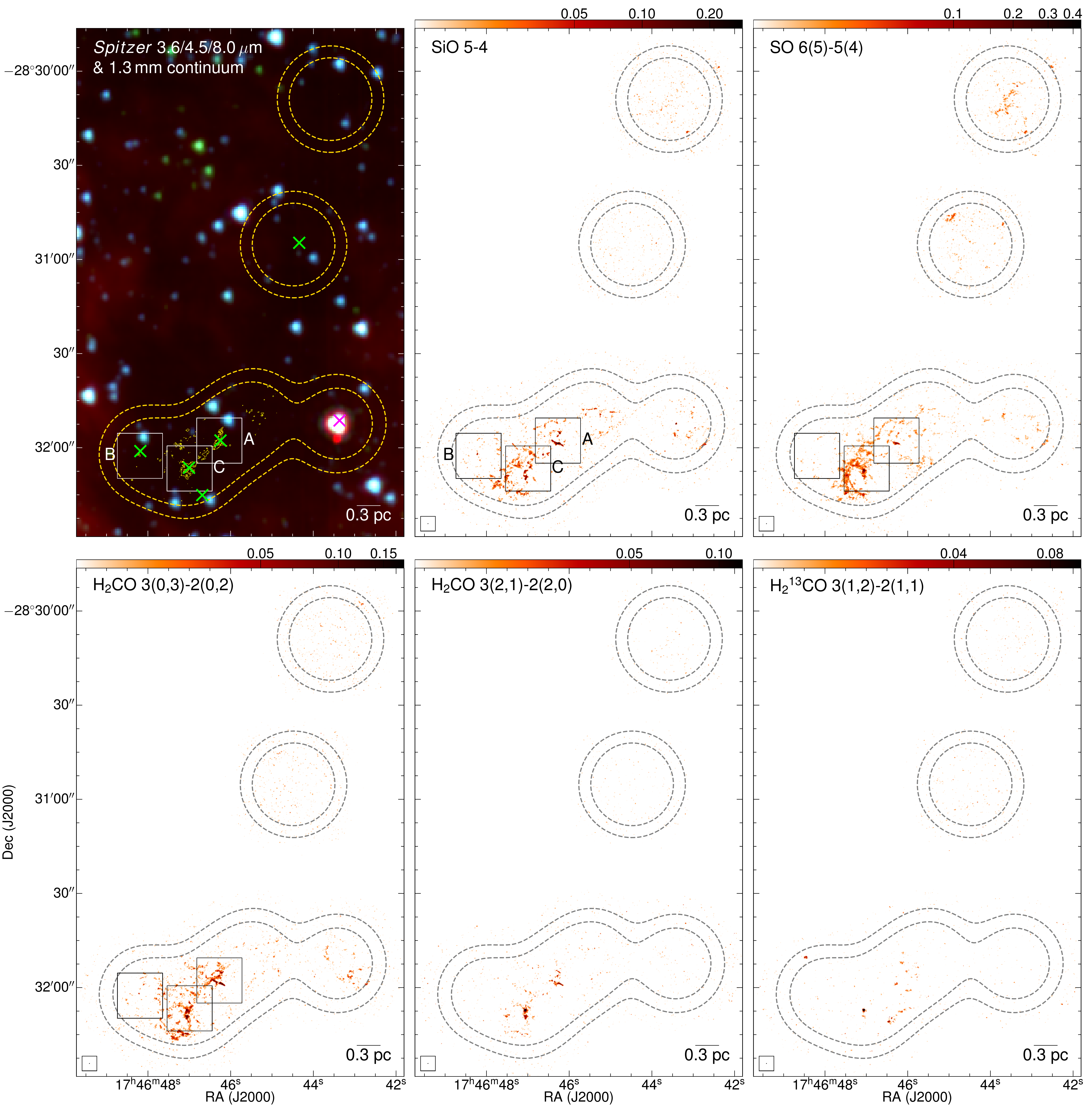}
\caption{Molecular line emission in \sgb{}. The layout of panels and symbols is the same as in \autoref{fig:c20kms}.}
\label{fig:sgrb1off}
\end{figure*}

\addtocounter{figure}{-1}
\begin{figure*}[!p]
\centering
\includegraphics[width=0.95\textwidth]{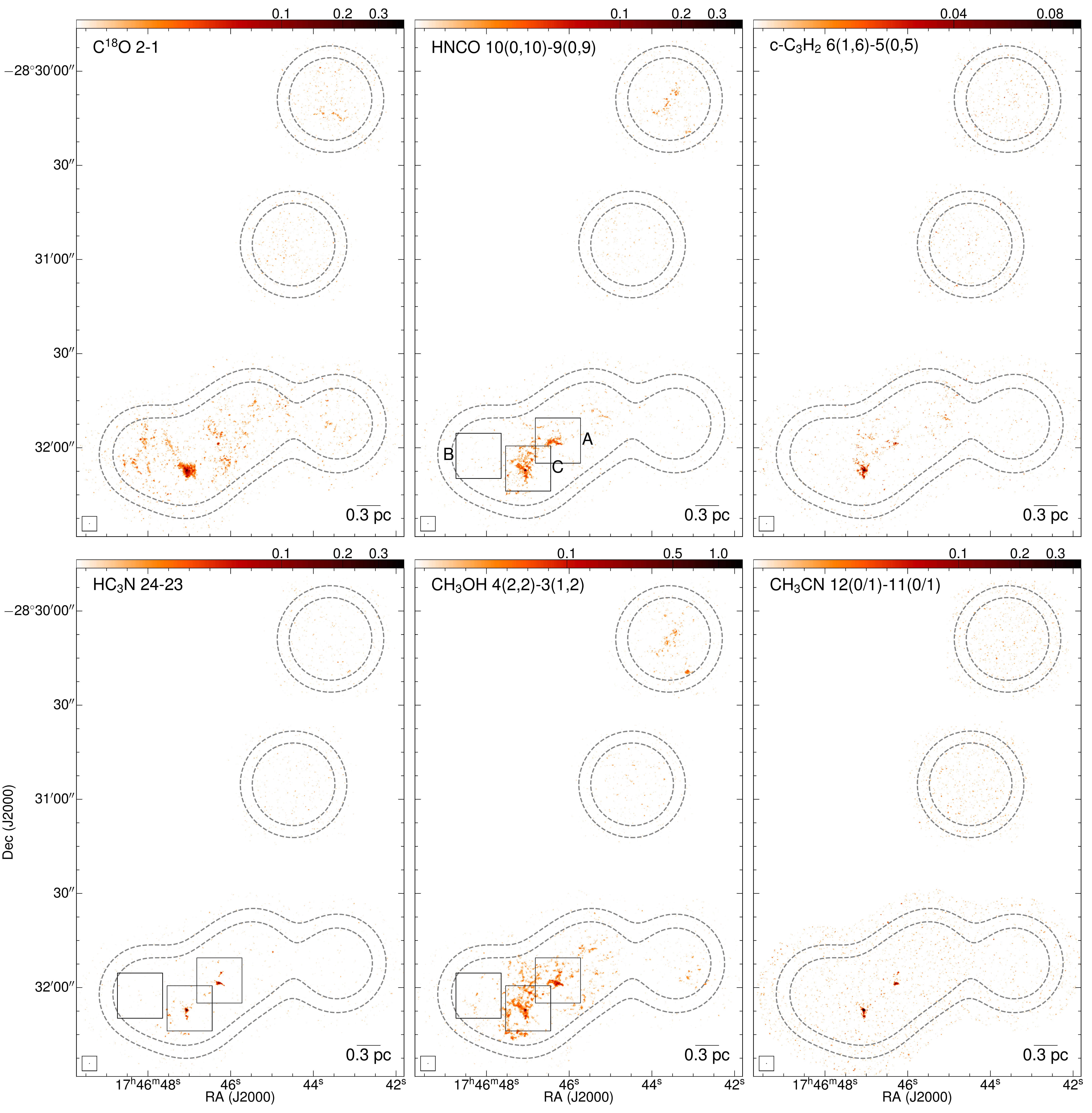}
\caption{Continued.}
\end{figure*}

\begin{figure*}[!p]
\centering
\includegraphics[width=0.95\textwidth]{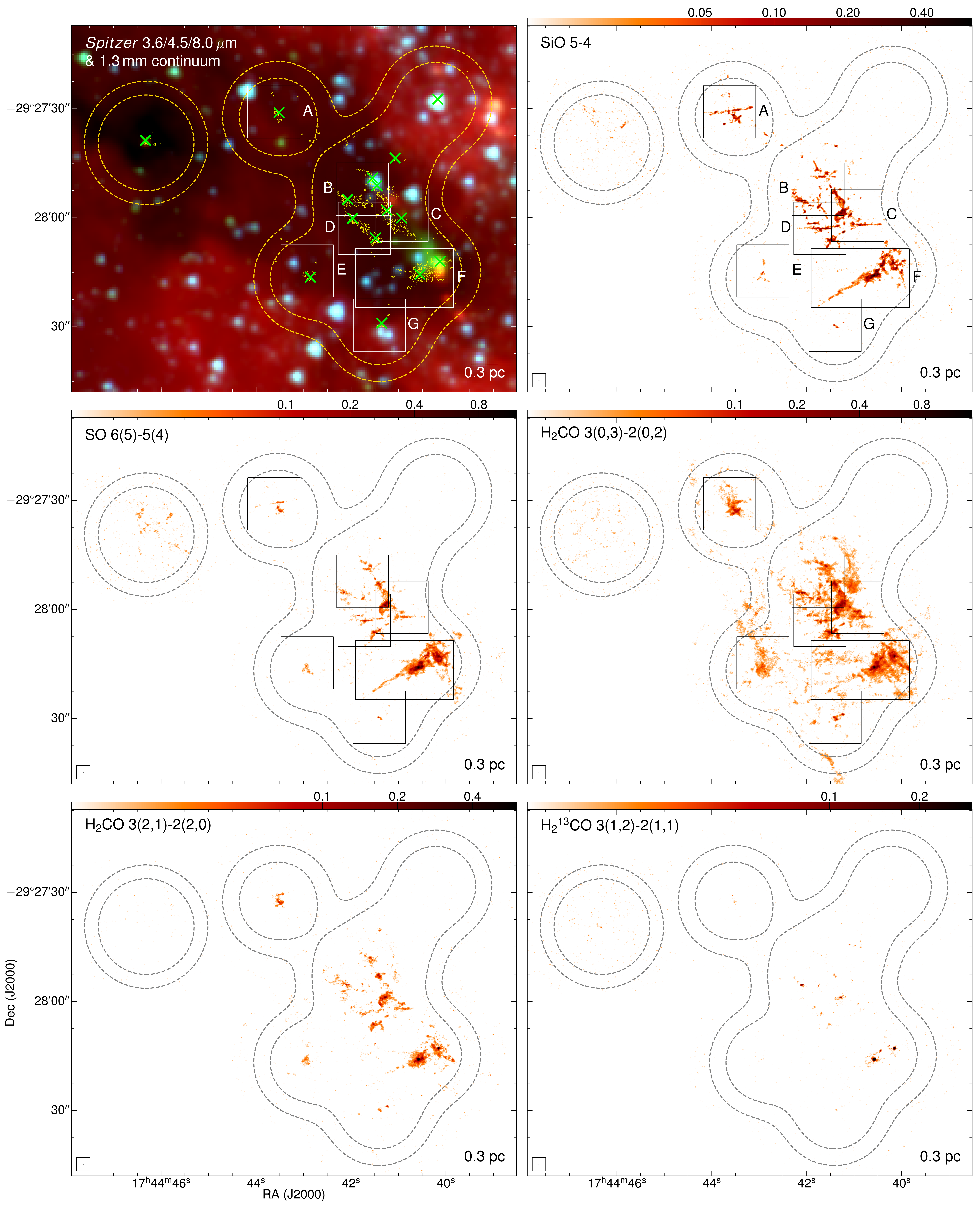}
\caption{Molecular line emission in Sgr~C. The layout of panels and symbols is the same as in \autoref{fig:c20kms}.}
\label{fig:sgrc}
\end{figure*}

\addtocounter{figure}{-1}
\begin{figure*}[!p]
\centering
\includegraphics[width=0.95\textwidth]{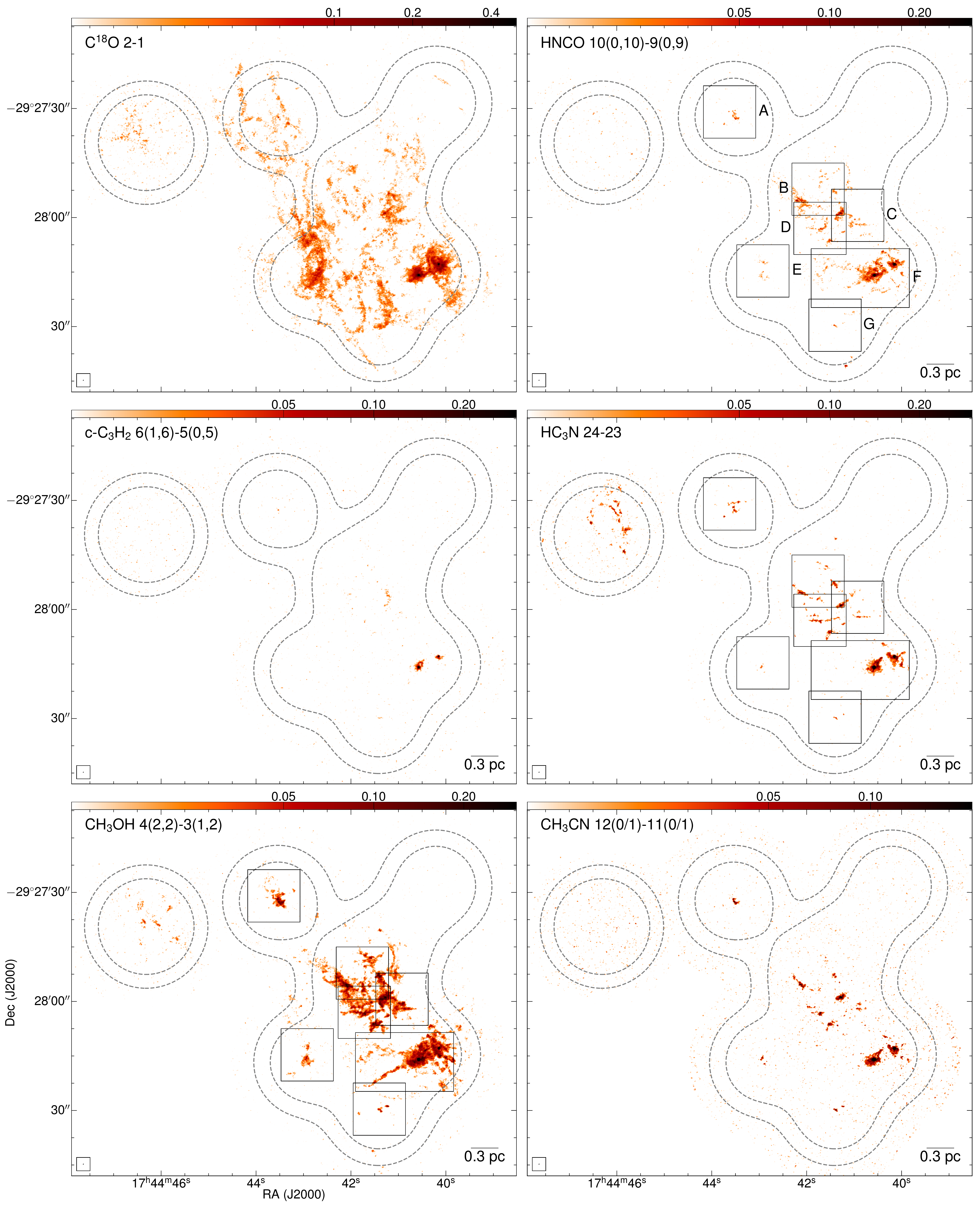}
\caption{Continued.}
\end{figure*}

\section{RESULTS}\label{sec:results}

\subsection{Overview of the Line Emission}\label{subsec:results_lines}
Typical spectra toward a chemically active core and a relatively quiescent region spatially offset from any cores but with line emission are shown in \autoref{fig:spec}. The former shows characteristic hot molecular core chemistry, while the latter represents regions that are likely influenced by pc scale shocks prevailing in the CMZ. Note that there exist even more chemically active cores (e.g., the two UC \hii{} regions in Sgr~C), with many more spectral lines detected, mostly from complex organic molecules. Here, we focus on the spatially extended spectral line emission detected outside of the hot cores or UC \hii{} regions, and leave the discussion of the line-rich chemically active cores to a future paper.
 
We identified spatially extended line species, and plotted their integrated intensities in Figures~\ref{fig:c20kms}--\ref{fig:sgrc}. The line species include the CO isotopologue \ceo{}, a group of potential shock tracers (SiO, SO, HNCO, \meth{}), and several dense gas tracers that are sometimes found in outflows (\fmh{}, \httco{}, \mthc{}, c-C$_3$H$_2$, \cyacet{}). \tco{} emission is more spatially extended than \ceo{} and is not plotted. Three features are clearly seen: i) Linear structures spatially associated with dust emission are prominent, which may be outflow lobes (black boxes in the figures). ii) Multiple lines are tracing similar filamentary structures that are spatially offset from any dust emission, including some that are more typically confined to hot cores in environments outside of the CMZ (e.g., \mthc{}), whose nature is unclear. An example is marked by the blue arrow in \autoref{fig:c20kms}. iii) Point-like SiO emission with large linewidths ($>$20~\kms{}) is found toward two \water{} masers that have known AGB star counterparts (magenta crosses in Figures~\ref{fig:c20kms} \& \ref{fig:sgrb1off}; see \citealt{lu2019a}), with no associated dust emission within a radius of 0.1~pc, which is probably originated from the atmosphere of AGB stars \citep{delgado2003}. Here we focus on the first feature, potential outflows, while leaving the discussion of the other features to future papers.

The two CO isotopologue lines, \tco{} and \ceo{}, present strong absorption at velocities of $-$55, $-$30, and $-$5~\kms{} against strong continuum emission (see the inset in \autoref{fig:spec}). These are consistent with the absorption features seen in other line observations toward the Galactic Center \citep[e.g.,][]{jones2012}, and are attributed to foreground gas in spiral arms along the line of sight. In particular, the absorption at $-$55~\kms{} is close to the cloud velocity of Sgr~C, which complicates the interpretation of the two lines in Sgr~C. Many of the lines, including \fmh{}, \meth{}, SiO, and the two CO isotopologues themselves, also present absorption features at a few \kms{} blue-shifted with respect to the cloud velocity toward continuum emission peaks (e.g., the absorption at $\sim$12~\kms{} in the inset of \autoref{fig:spec}). These features are likely owing to a combination of missing flux as a result of interferometer observations and self absorption when the lines become optically thick.

\subsection{Identification of Outflows}\label{subsec:results_outflows}
Signatures of protostellar outflows have been detected toward Sgr~B2(M) and (N), the two high-mass protoclusters in the Sgr~B2 complex \citep{qin2008,higuchi2015,bonfand2017}. Outside of Sgr~B2, detecting protostellar outflows has been challenging in the CMZ, because of the lack of resolution to spatially resolve outflows and the prevalence of broad linewidth gas produced by phenomena other than outflows \citep{henshaw2019,sormani2019}. Previous Submillimeter Array (SMA) observations have detected widespread emission of potential shock tracers (e.g., SiO, SO, \meth{}) at 0.2~pc scales in CMZ clouds \citep{kauffmann2013a,lu2017,cmzoom2020}. However, limited by the angular resolution and the imaging sensitivity, it was unclear whether the emission seen by the SMA is owing to protostellar outflows or pc-scale shocks prevailing in the CMZ. Only recently, ALMA observations with high resolution and high sensitivity start to detect outflows in the CMZ, e.g., in the G0.253$+$0.025 cloud (D.\ Walker et al.\ submitted, 2020).

Our high angular resolution ALMA observations resolved substructures of 2000 AU scale in the molecular line emission, enabling us to search for protostellar outflows. The integrated intensity maps in Figures~\ref{fig:c20kms}--\ref{fig:sgrc} already reveal collimated emission spatially associated with cores, indicating the existence of outflows. 
 
We then examined potential shock tracers detected by ALMA, including SiO, SO, \fmh{}, HNCO, \cyacet{}, and \meth{}. All these tracers have been previously found to be enhanced by at least one order of magnitude in shocked regions in protostellar outflows \citep[e.g.,][]{bachiller1997}. We applied the following criteria to identify outflows: 

i) We used the \water{} masers from \citet{lu2019a} as a guidance to search for associated shock tracer emission. First, we  made integrated intensity maps of SiO across the full velocity range, and searched for linear structures spatially associated with the masers. If linear emission is found, then, we made integrated intensity maps of blue and red shifted components based on the \vlsr{} of the cloud (see \autoref{tab:clouds}), and checked whether the linear structures show symmetric blue and red shifted emission with respective to the masers. Finally, we determined the systematic velocity of each individual outflow driving source, by using dense gas tracers in the ALMA data toward the maser position (\mthc{}, \cyacet{}, \meth{}, or \ceo{}, in a decreasing order of preference; \ceo{} was used only once, towards \ctw{}-F \#1), and remade integrated intensity maps of blue and red shifted components based on the individual \vlsr{}. If the shock tracer emission exhibits blue and red shifted components with respect to the systematic velocity at opposite positions to the maser position, it is considered as an outflow.

ii) In cases where \water{} masers are not present, we checked the emission of the six tracers around the 2000-AU scale cores from \citetalias{lu2020a} following the same procedures, to search for blue and/or red shifted line emission spatially offset from the cores. We require the blue and/or red shifted features to be seen in at least two shock tracers, including the canonical shock tracer SiO, plus any of the five supplemental tracers.

\begin{figure*}[!t]
\centering
\includegraphics[width=0.86\textwidth]{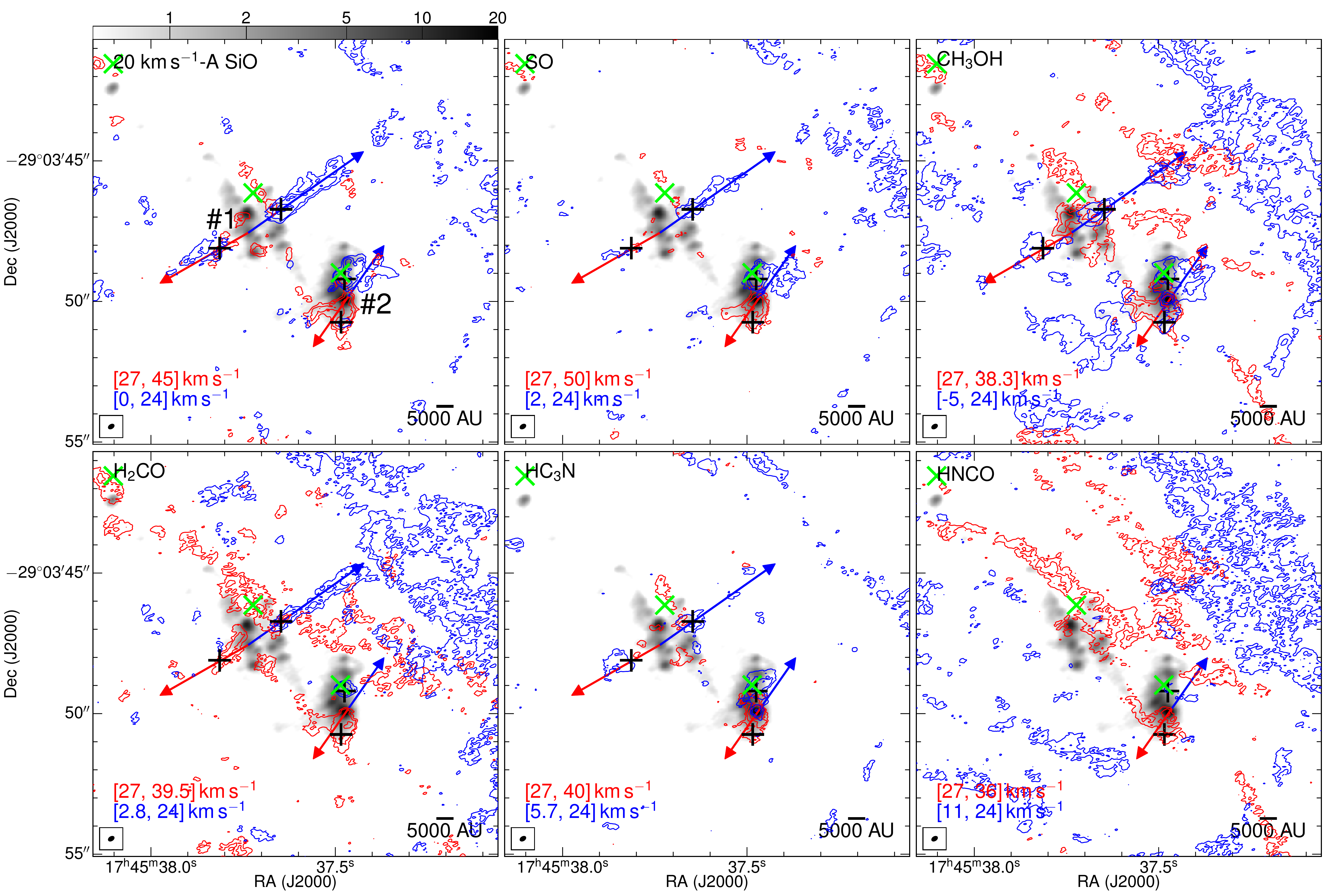}
\caption{Outflows in region A in \ctw{}. The gray scale image in all panels shows the ALMA 1.3~mm continuum emission, with the scale bar at the top in unit of \mjypbm{} in a logarithmic scale. The green crosses mark positions of the \water{} masers in \citet{lu2019a}. In each panel, the blue and red contours illustrate the blue and red shifted line emission integrated within the specified velocity ranges, at levels of [3,6,12]$\sigma$, where $\sigma=\sqrt{N}\sigma_cv_c$, $N$ being the number of channel, $\sigma_c$ the rms of individual channels, and $v_c$ the channel width in \kms{}. When there are multiple outflows in the panel, the velocity ranges are chosen to highlight the bipolar morphologies of all the outflows, but for individual outflows, the blue and red shifted lobes may extend beyond these velocity ranges or be contaminated by diffuse gas, which can be better visualized in \autoref{fig:pv_sio}. Identified outflows are highlighted by blue and red arrows, but note that some outflows are not seen in all the lines (in the case shown here, outflow \#1 is not detected in HNCO). The thick black crosses mark the reference positions we choose to derive column densities of the molecules based on which we estimate molecular abundances.}
\label{fig:outflows_c20_a}
\end{figure*}

\begin{figure*}[!t]
\centering
\includegraphics[width=0.86\textwidth]{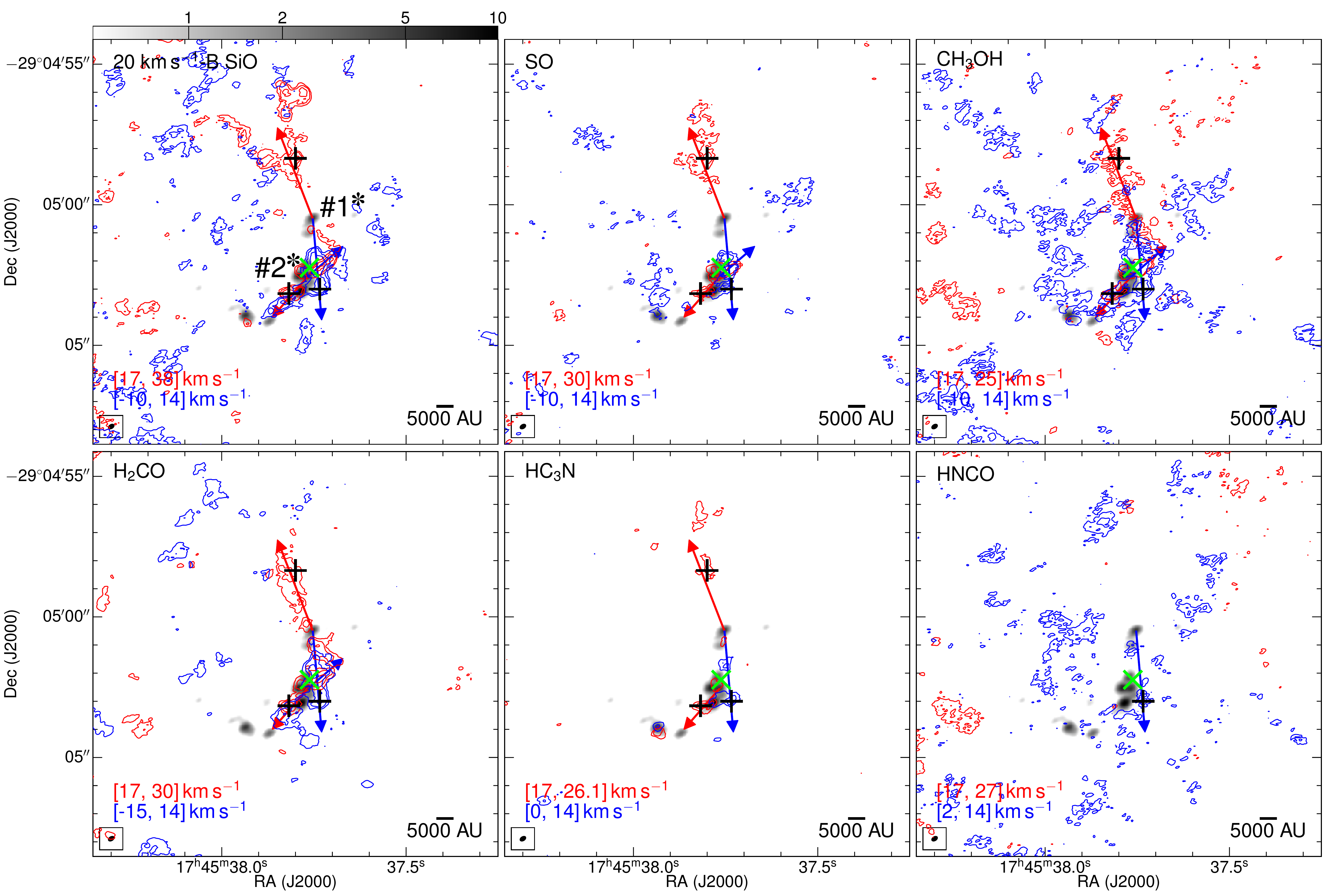}
\caption{Outflows in region B in \ctw{}.}
\label{fig:outflows_c20_b}
\end{figure*}

\begin{figure*}[!t]
\centering
\includegraphics[width=0.86\textwidth]{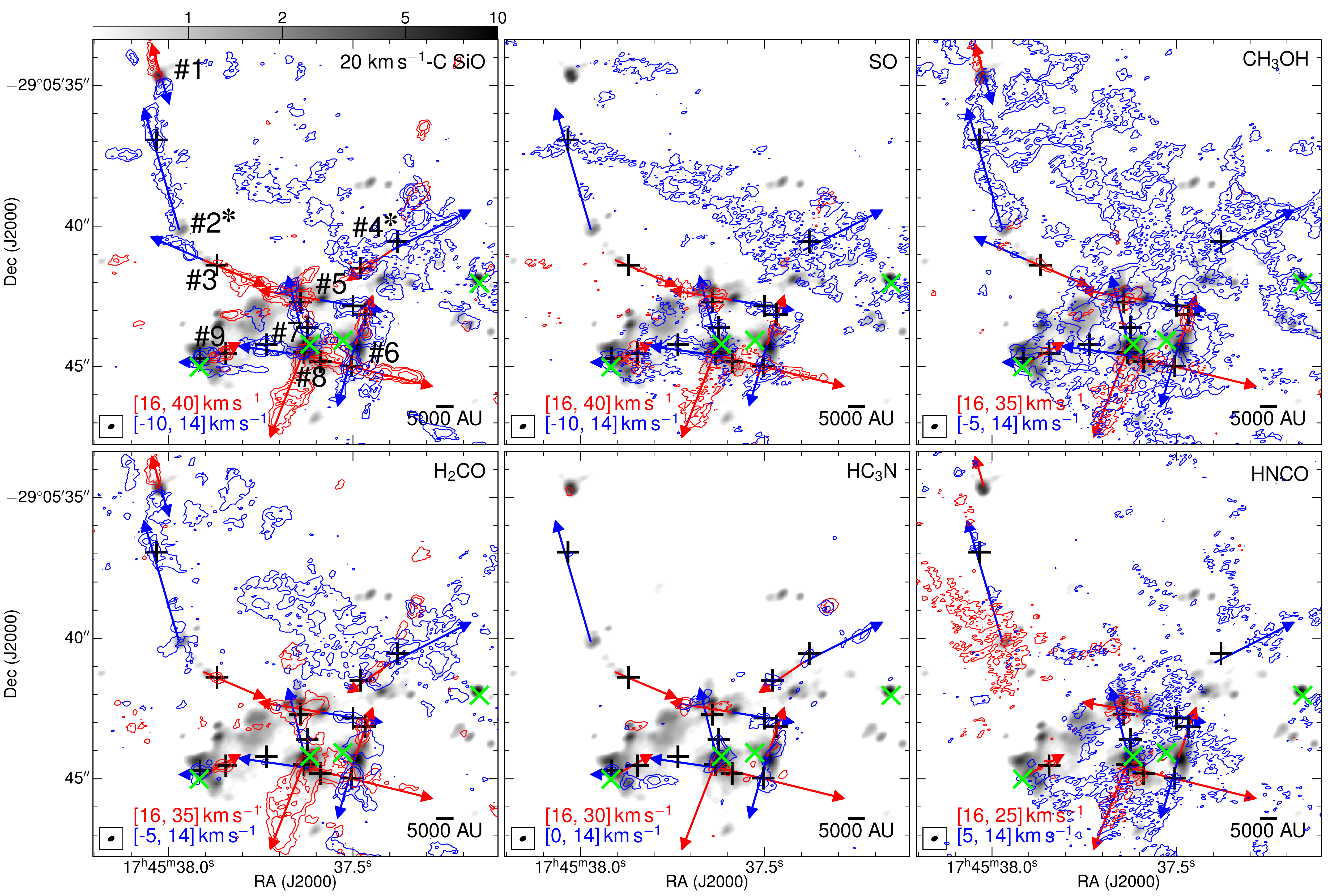}
\caption{Outflows in region C in \ctw{}.}
\label{fig:outflows_c20_c}
\end{figure*}

\begin{figure*}[!t]
\centering
\includegraphics[width=0.86\textwidth]{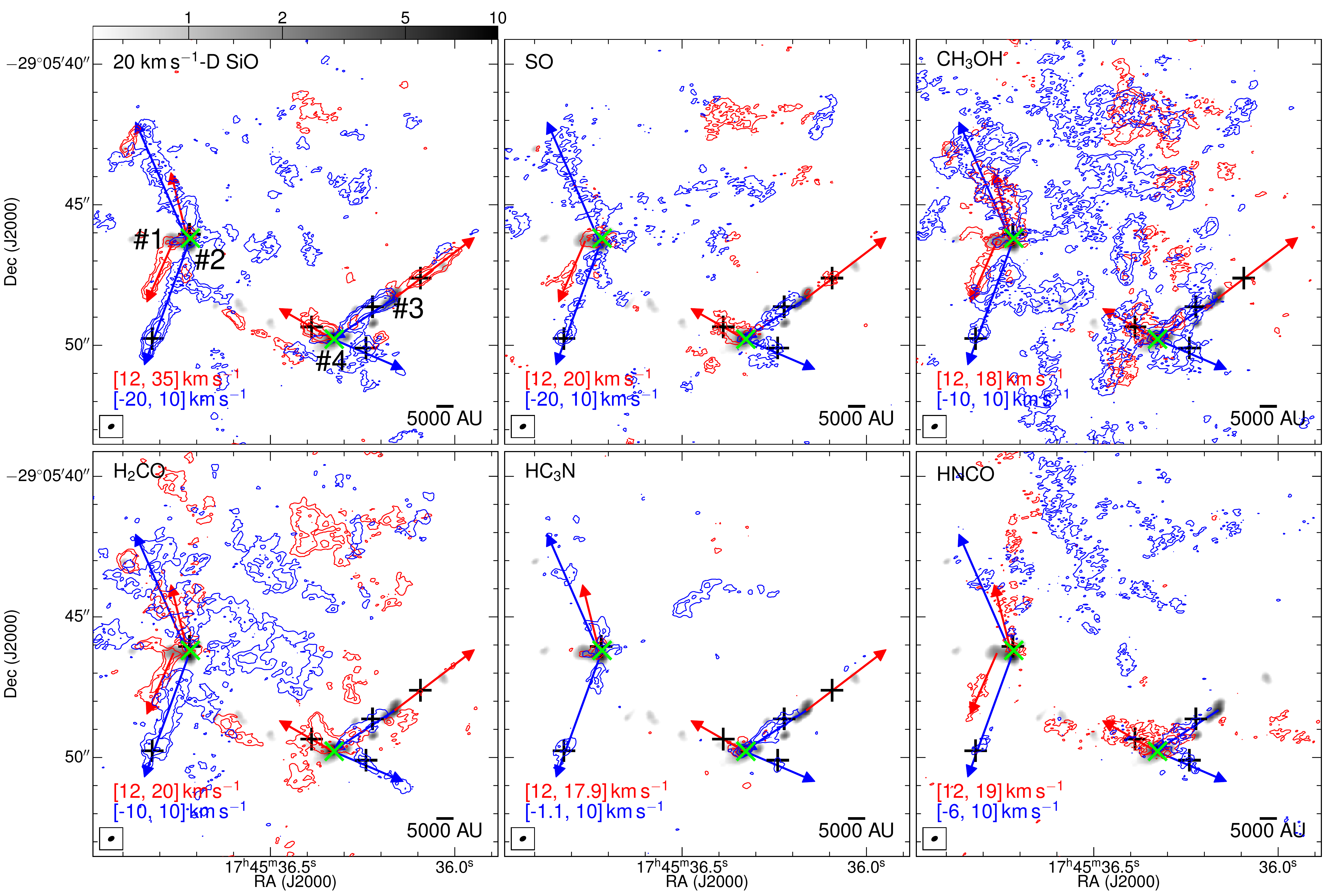}
\caption{Outflows in region D in \ctw{}.}
\label{fig:outflows_c20_d}
\end{figure*}

\begin{figure*}[!t]
\centering
\includegraphics[width=0.86\textwidth]{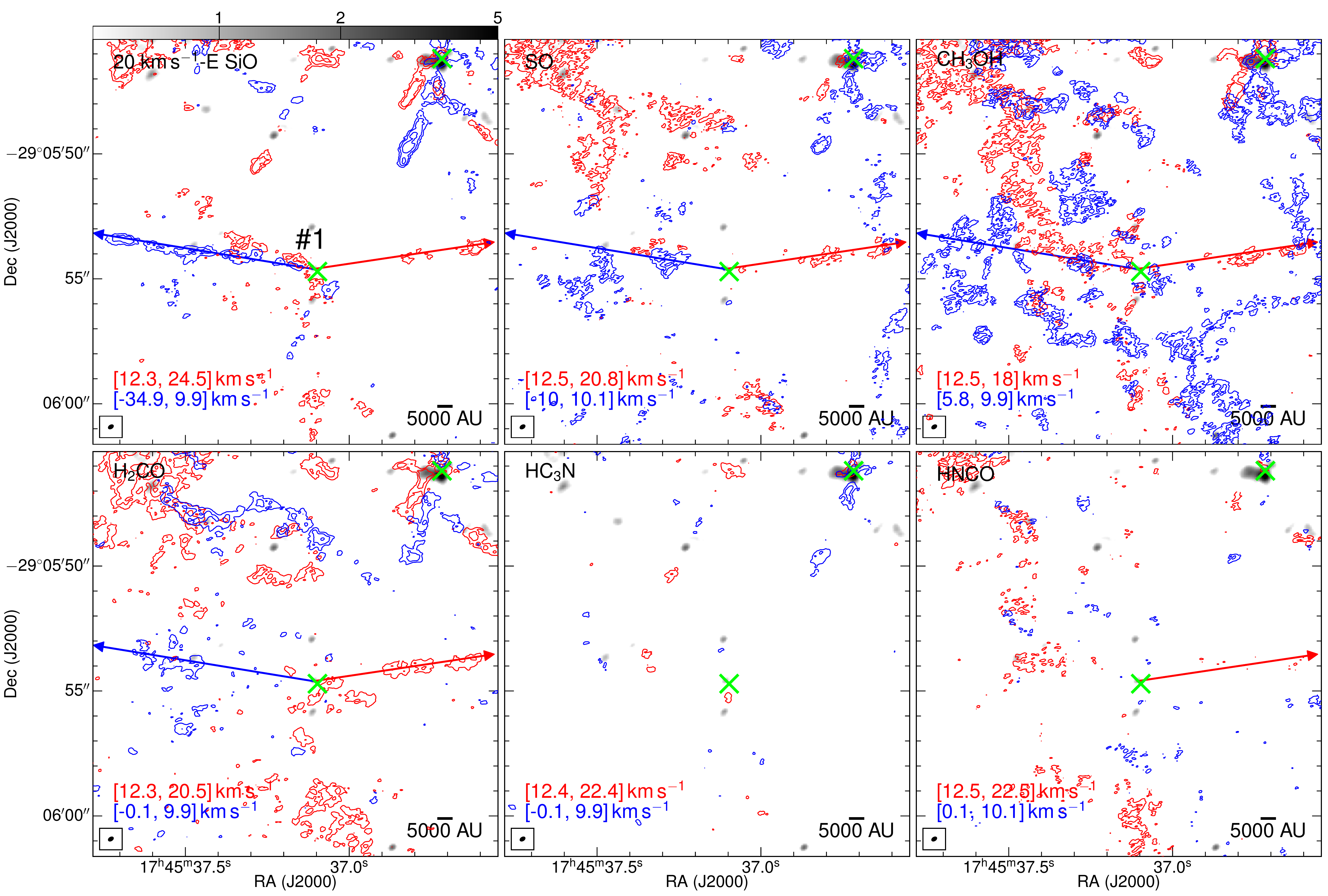}
\caption{Outflows in region E in \ctw{}.}
\label{fig:outflows_c20_e}
\end{figure*}

\begin{figure*}[!t]
\centering
\includegraphics[width=0.86\textwidth]{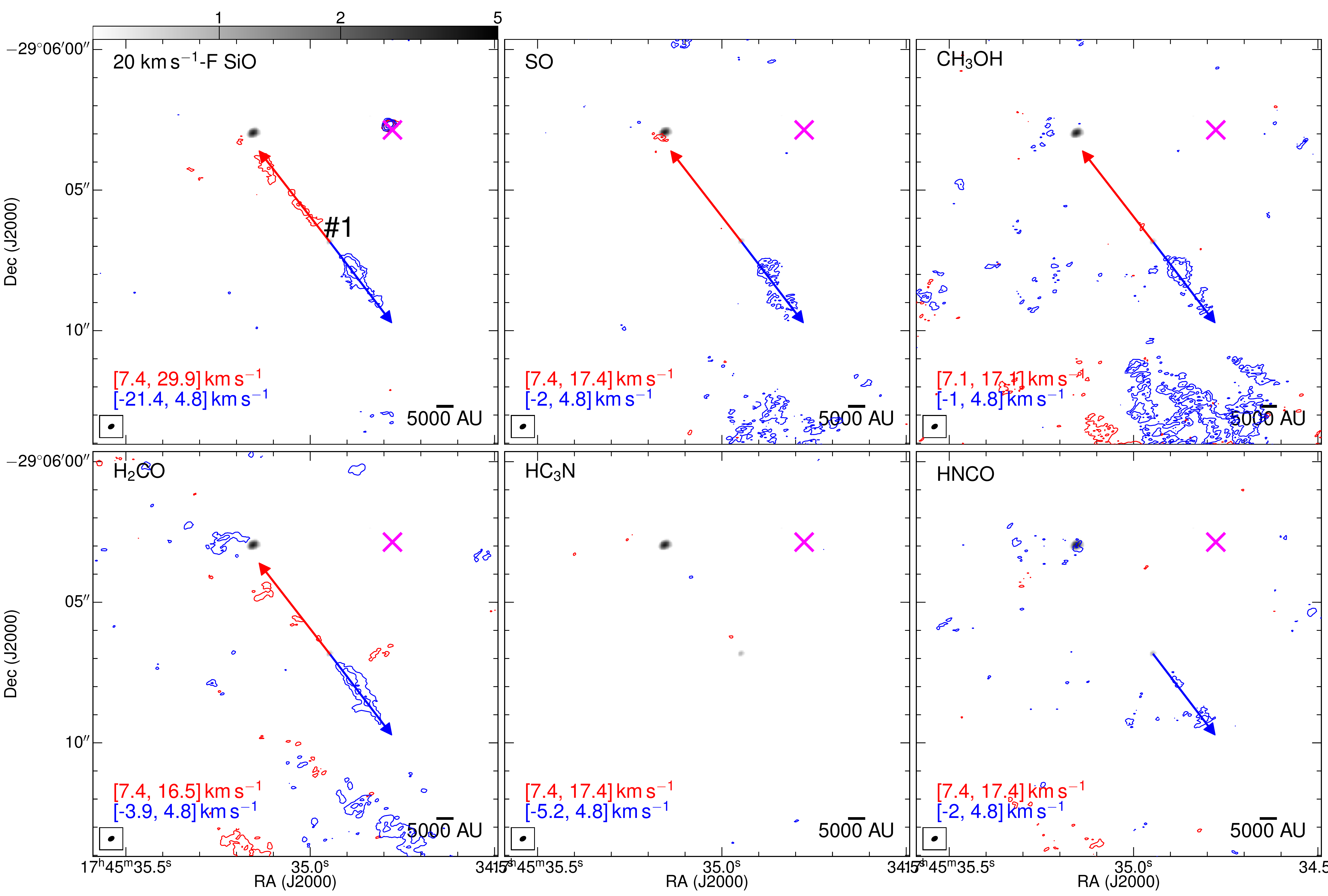}
\caption{Outflows in region F in \ctw{}.}
\label{fig:outflows_c20_f}
\end{figure*}

\begin{figure*}[!t]
\centering
\includegraphics[width=0.86\textwidth]{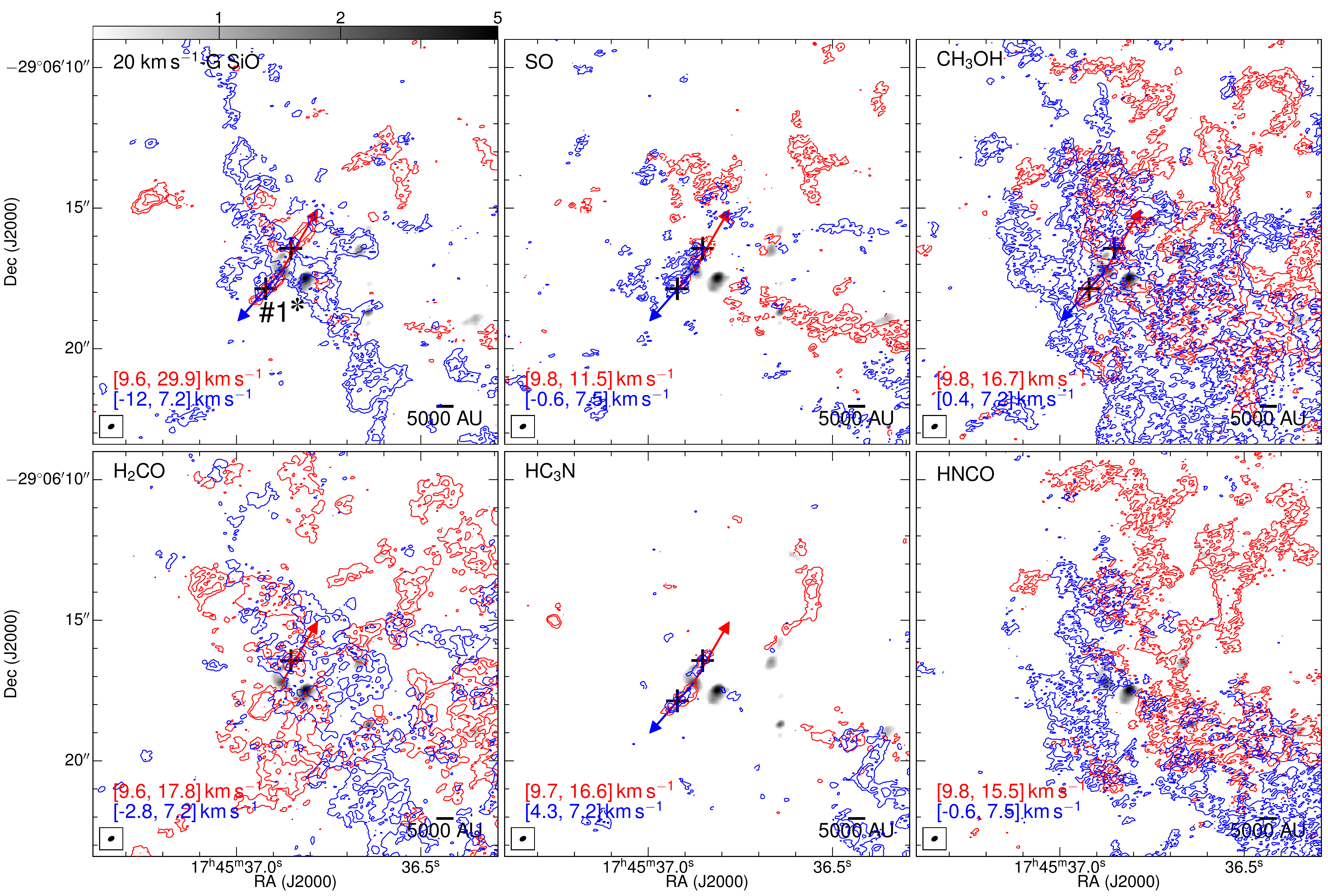}
\caption{Outflows in region G in \ctw{}.}
\label{fig:outflows_c20_g}
\end{figure*}

\begin{figure*}[!t]
\centering
\includegraphics[width=0.86\textwidth]{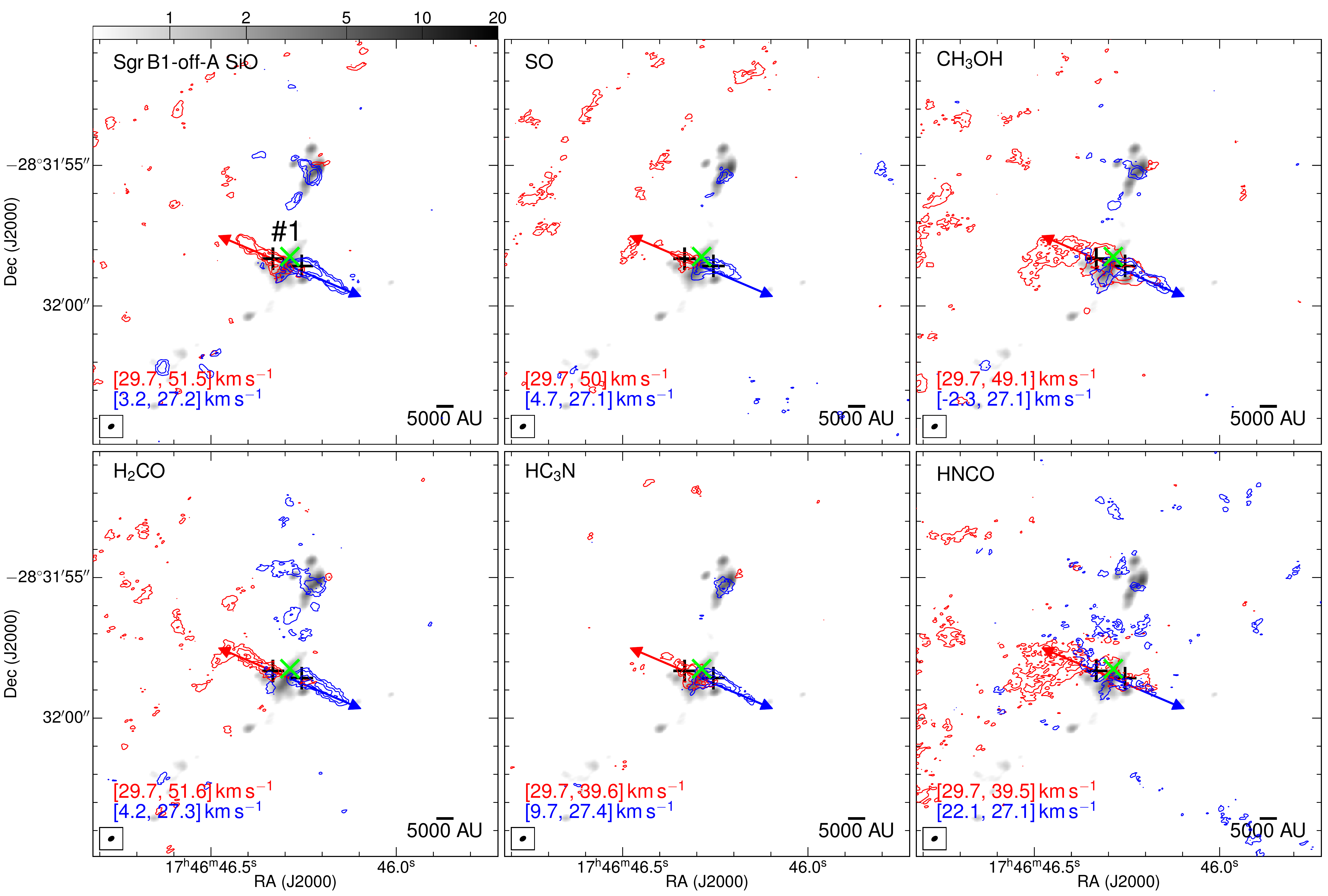}
\caption{Outflows in region A in \sgb{}.}
\label{fig:outflows_sgb_a}
\end{figure*}

\begin{figure*}[!t]
\centering
\includegraphics[width=0.86\textwidth]{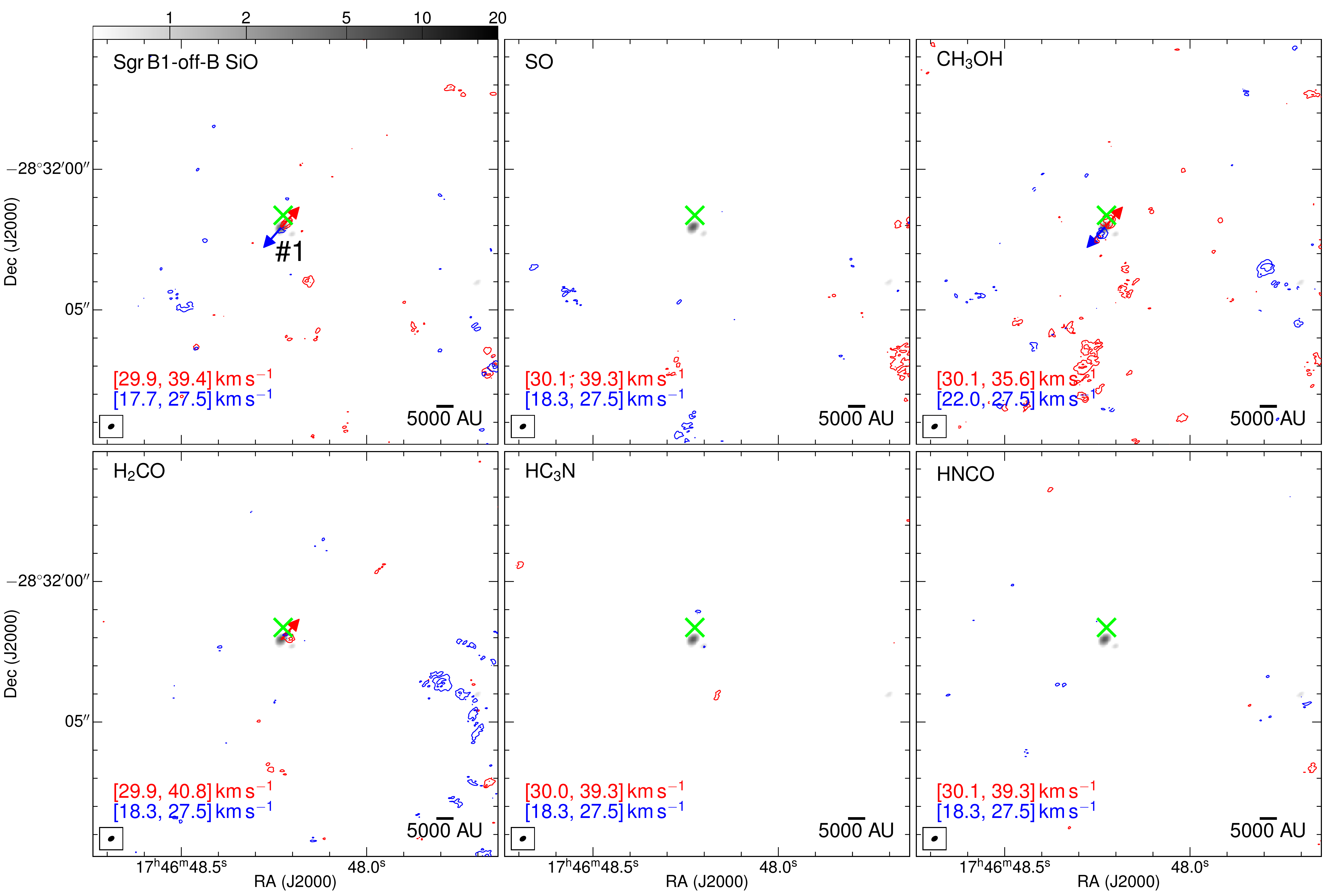}
\caption{Outflows in region B in \sgb{}.}
\label{fig:outflows_sgb_b}
\end{figure*}

\begin{figure*}[!t]
\centering
\includegraphics[width=0.86\textwidth]{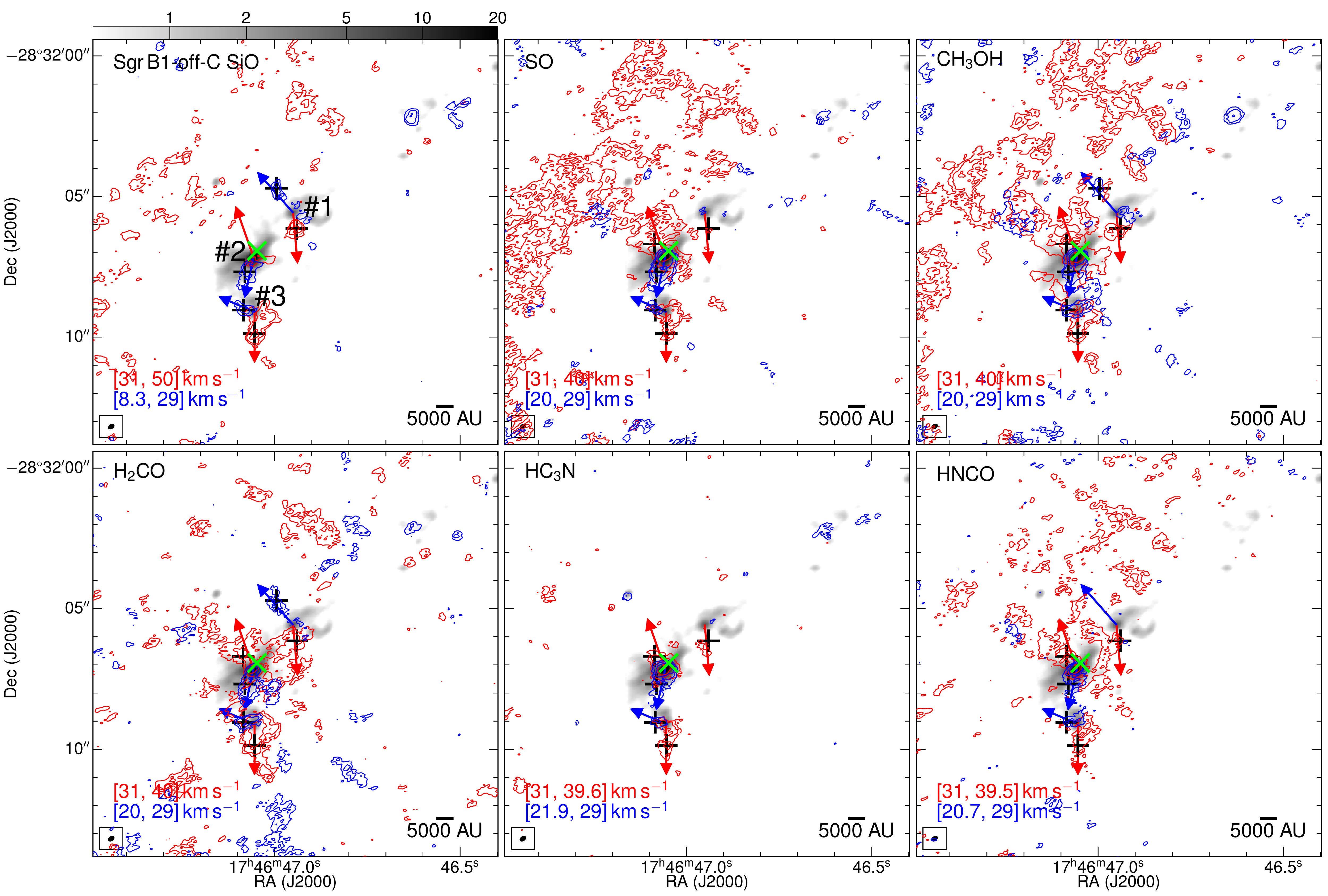}
\caption{Outflows in region C in \sgb{}.}
\label{fig:outflows_sgb_c}
\end{figure*}

\begin{figure*}[!t]
\centering
\includegraphics[width=0.86\textwidth]{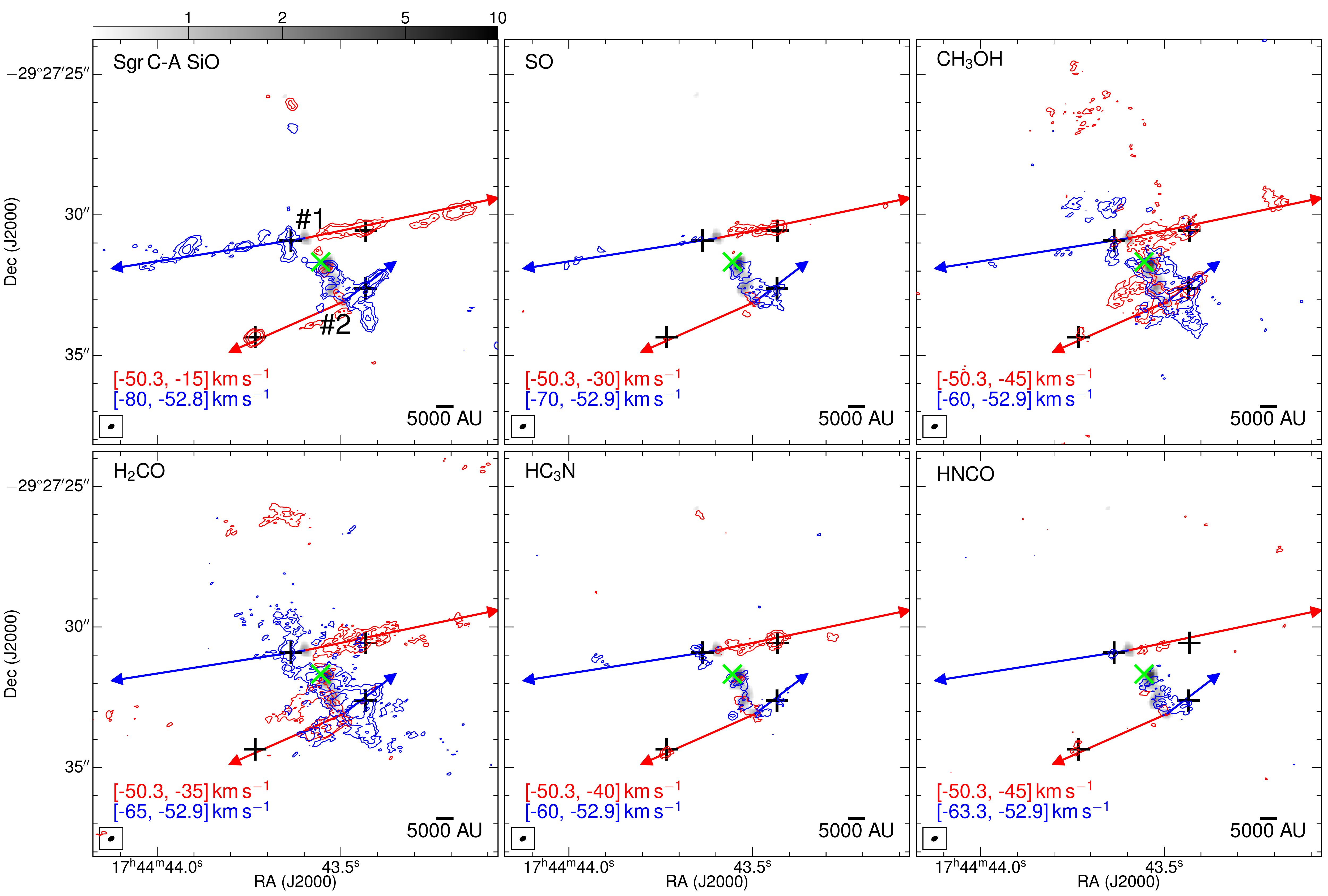}
\caption{Outflows in region A in Sgr~C.}
\label{fig:outflows_sgrc_a}
\end{figure*}

\begin{figure*}[!t]
\centering
\includegraphics[width=0.86\textwidth]{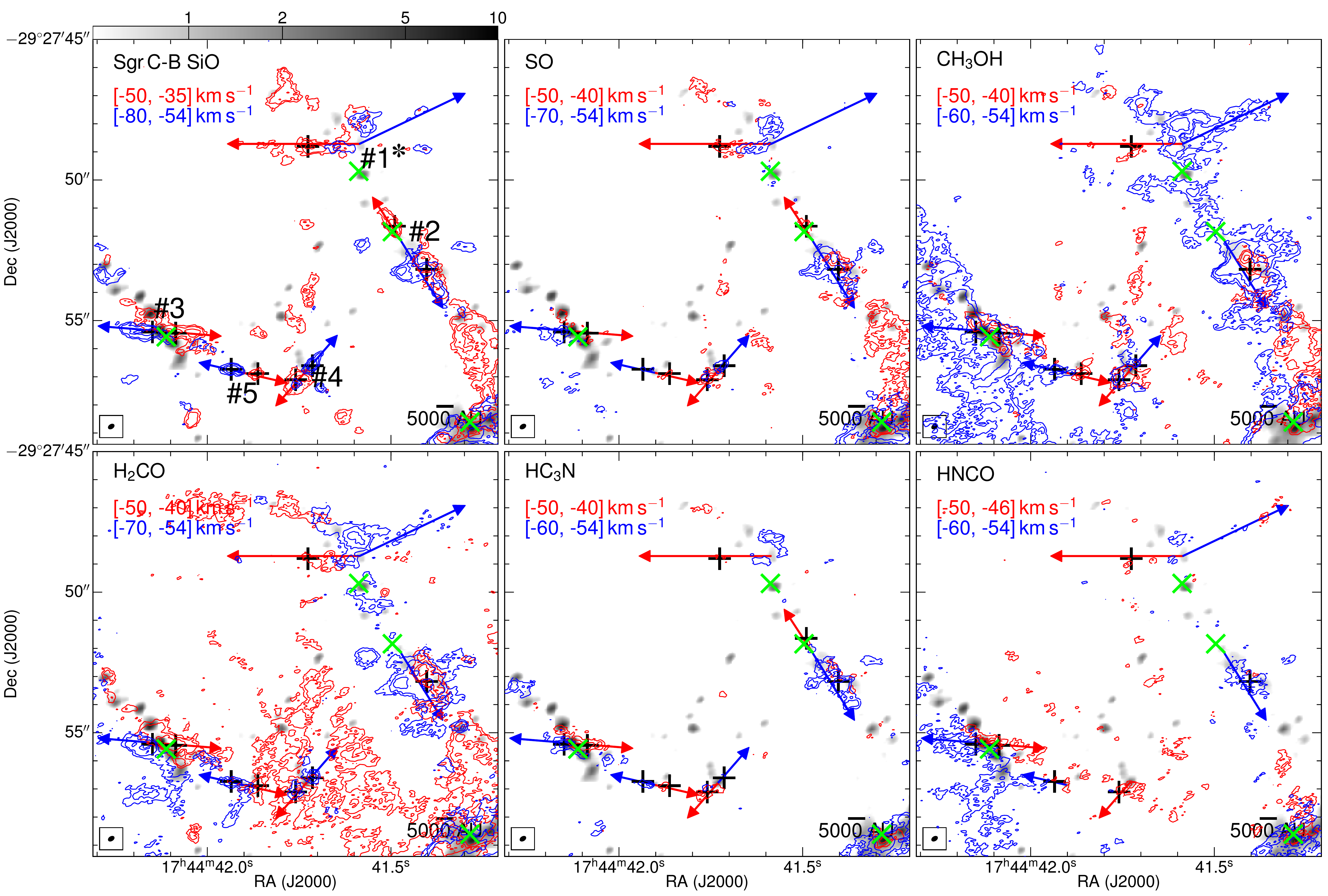}
\caption{Outflows in region B in Sgr~C.}
\label{fig:outflows_sgrc_b}
\end{figure*}

\begin{figure*}[!t]
\centering
\includegraphics[width=0.86\textwidth]{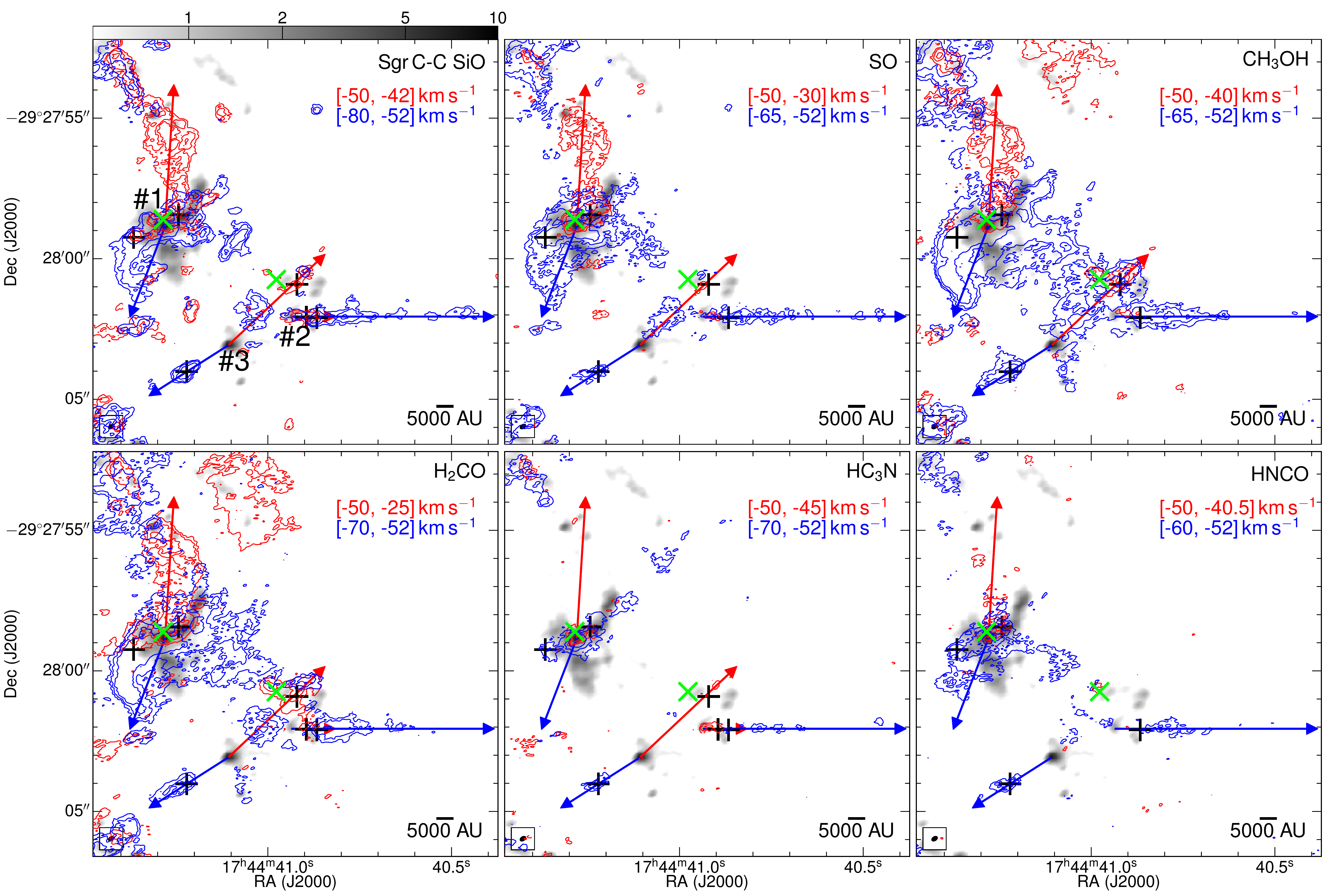}
\caption{Outflows in region C in Sgr~C.}
\label{fig:outflows_sgrc_c}
\end{figure*}

\begin{figure*}[!t]
\centering
\includegraphics[width=0.86\textwidth]{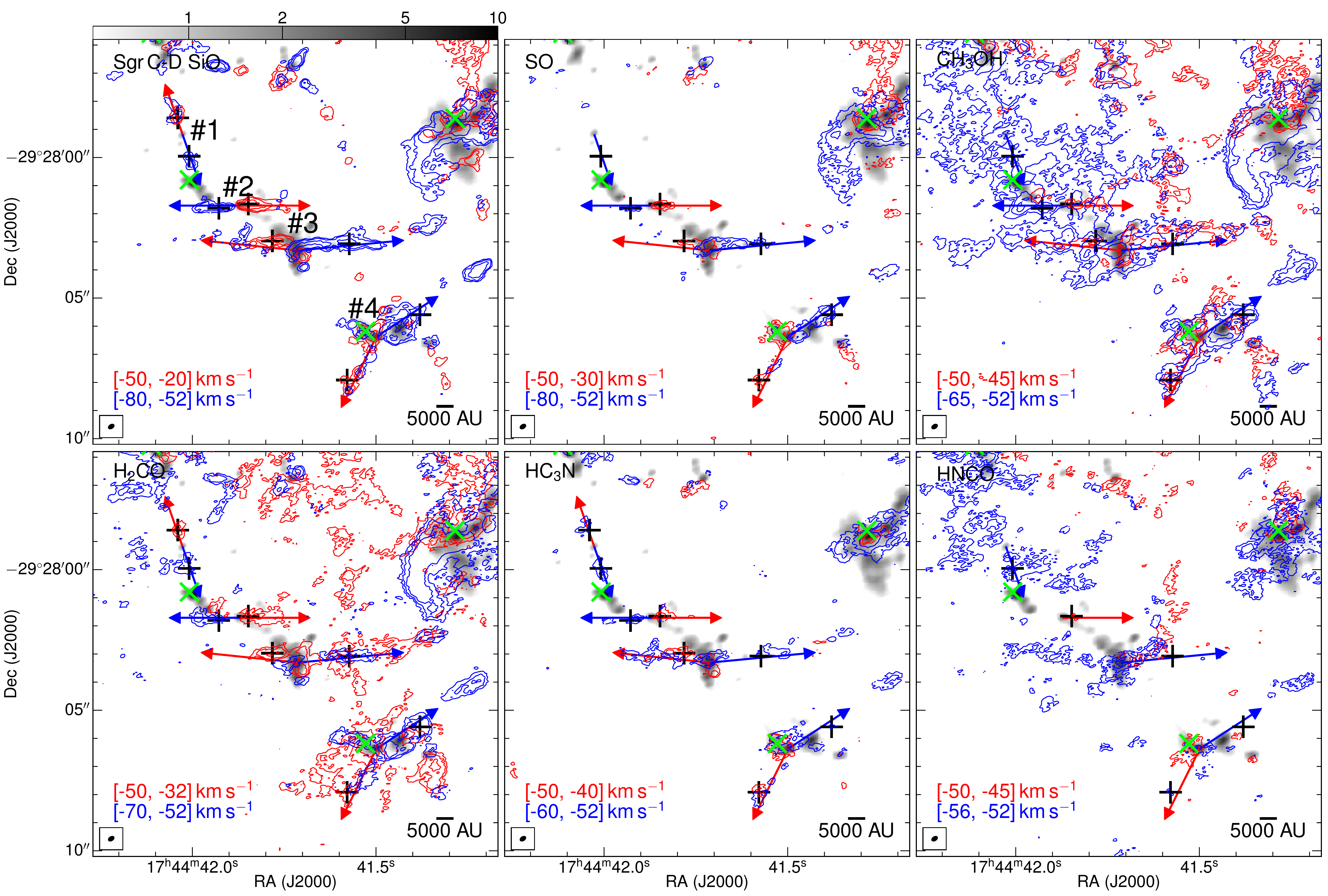}
\caption{Outflows in region D in Sgr~C.}
\label{fig:outflows_sgrc_d}
\end{figure*}

\begin{figure*}[!t]
\centering
\includegraphics[width=0.86\textwidth]{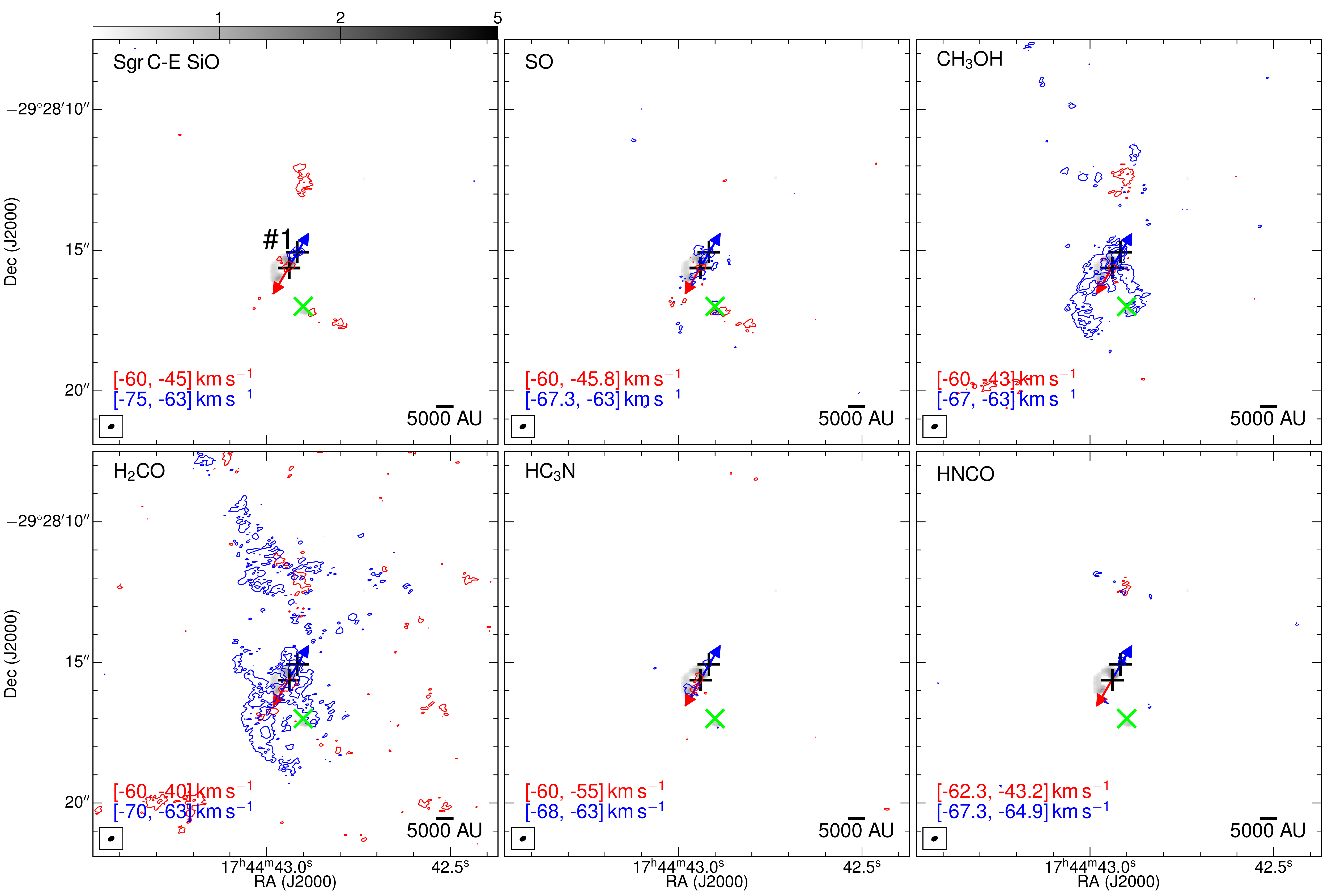}
\caption{Outflows in region E in Sgr~C.}
\label{fig:outflows_sgrc_e}
\end{figure*}

\begin{figure*}[!t]
\centering
\includegraphics[width=0.86\textwidth]{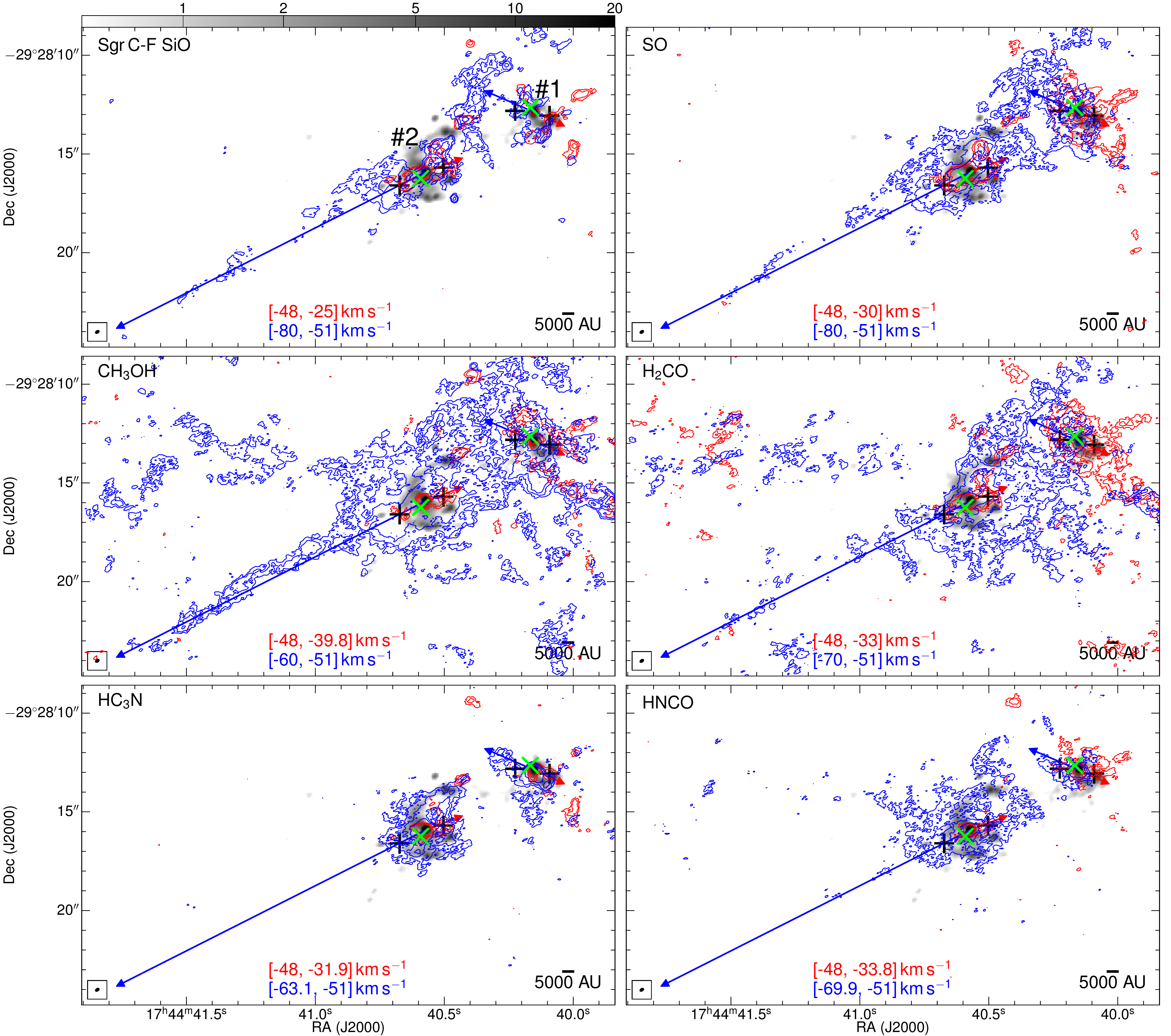}
\caption{Outflows in region F in Sgr~C.}
\label{fig:outflows_sgrc_f}
\end{figure*}

\begin{figure*}[!t]
\centering
\includegraphics[width=0.86\textwidth]{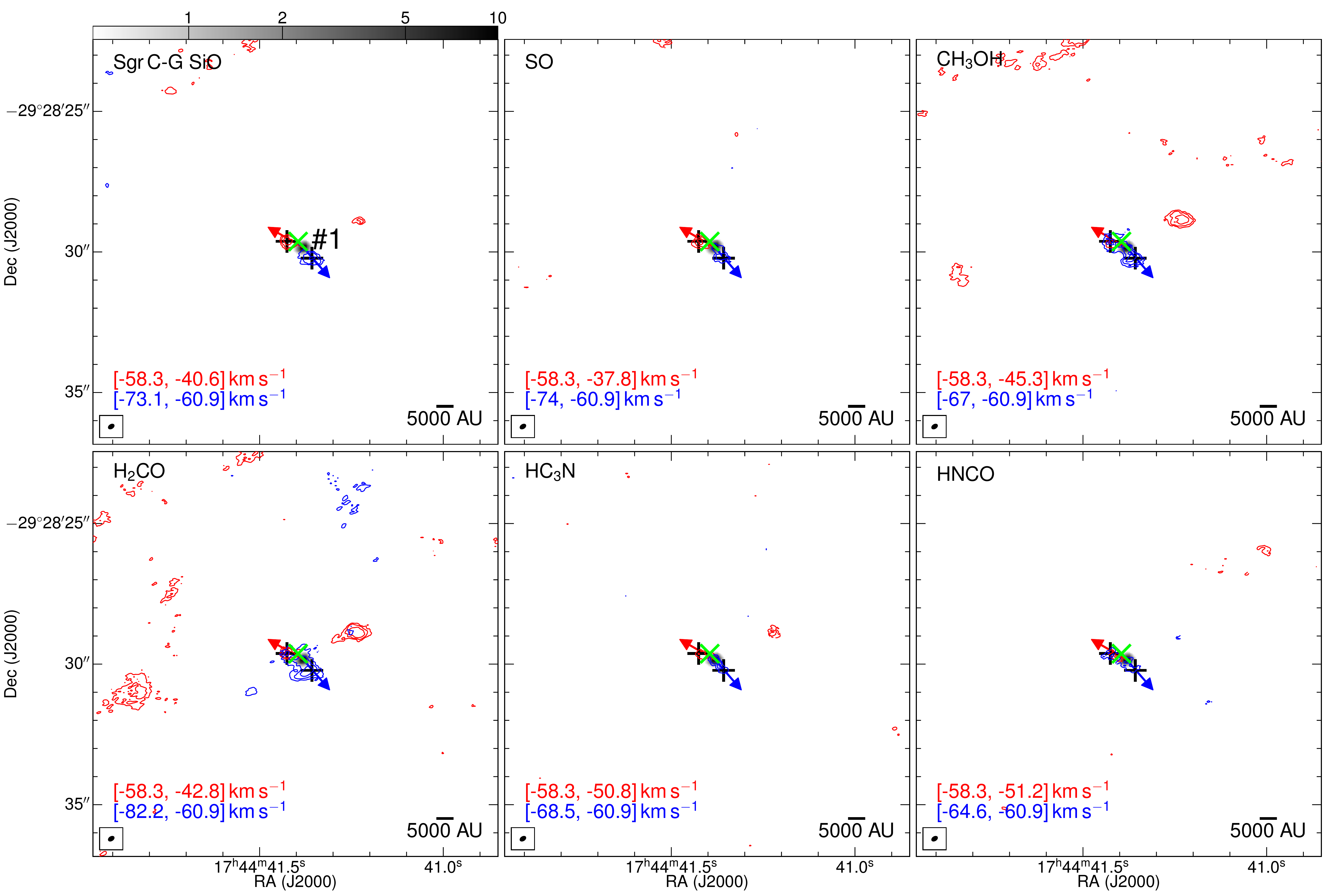}
\caption{Outflows in region G in Sgr~C.}
\label{fig:outflows_sgrc_g}
\end{figure*}

\begin{figure*}[!t]
\centering
\includegraphics[width=1.0\textwidth]{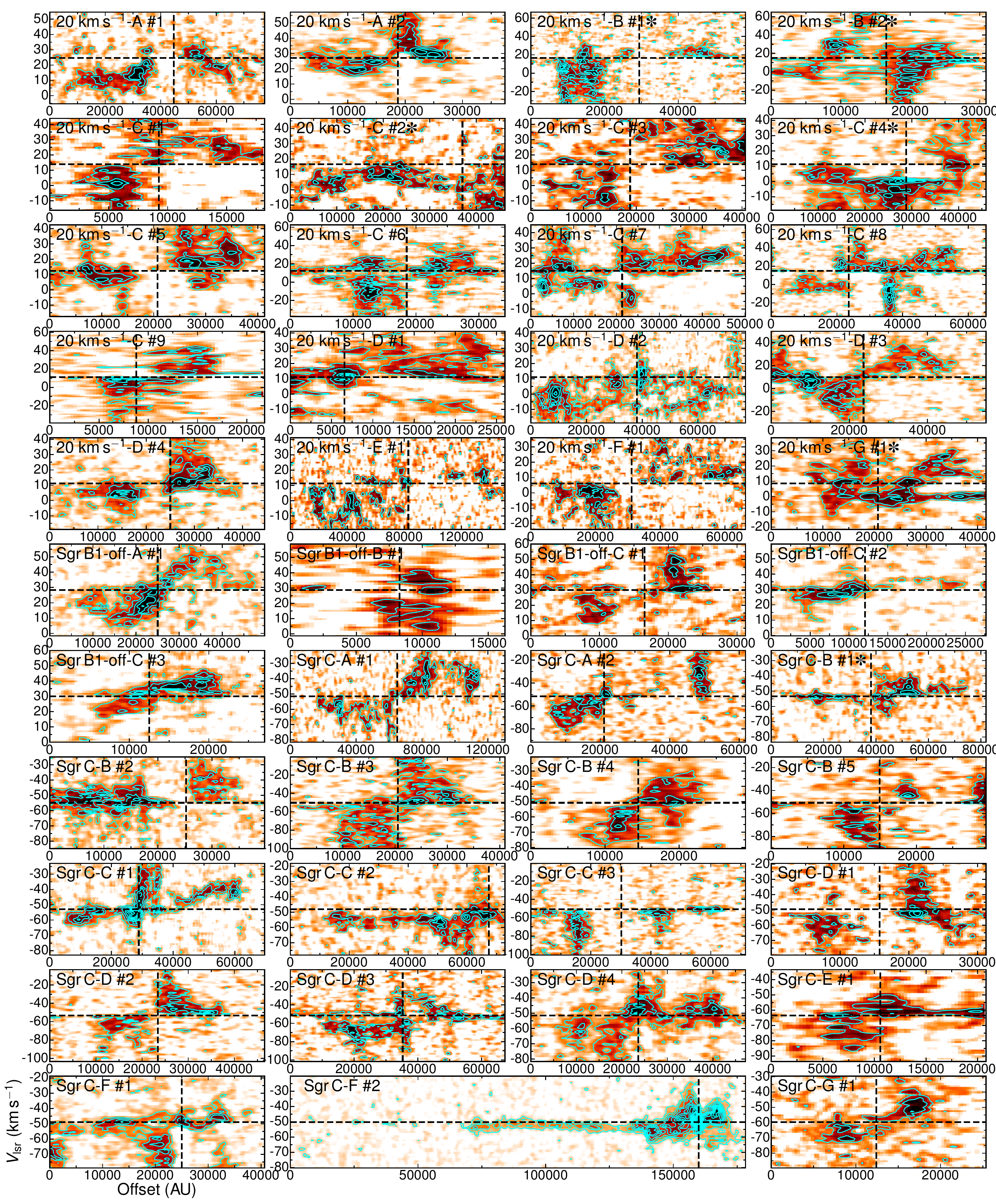}
\vspace{-2em}
\caption{Position-velocity diagrams of the 43 outflows using the SiO line. The slices are taken along the arrows shown in Figures~\ref{fig:outflows_c20_a}--\ref{fig:outflows_sgrc_g}, from the blue-shifted to red-shifted side, averaged within a width of 5000~AU. The background image and the cyan contours show the SiO emission, with contour levels starting at 2$\sigma$ in step of 2$\sigma$ where $\sigma$$\approx$1~\mjypbm{}. The x axis is the spatial offset along the slices, and the y axis is \vlsr{}. The vertical and horizontal dashed lines mark the position and \vlsr{} of the cores, respectively.}
\label{fig:pv_sio}
\end{figure*}

We identified 43 outflows, and marked regions where they are detected with boxes in Figures~\ref{fig:c20kms}--\ref{fig:sgrc}. The zoomed-in views are in Figures~\ref{fig:outflows_c20_a}--\ref{fig:outflows_sgrc_g}, in which we plot the red/blue shifted shock tracer emission with respect to the systematic velocity and highlight individual outflows with arrows. The position-velocity diagrams of the 43 outflows made from the SiO line are displayed in \autoref{fig:pv_sio}. The numbers of outflows identified in the individual clouds are listed in \autoref{tab:clouds}.

In several cases, lobes from different outflows spatially overlap with each other (e.g., outflows \#5--\#9 in region C of \ctw{}; see \autoref{fig:outflows_c20_c}), but we were able to separate them apart unambiguously based on velocities (see \autoref{fig:pv_sio}). Most of the other outflows are easily distinguished spatially from nearby outflows and the diffuse emission. All these outflows are considered as `highly likely'. However, there exist cases where the outflows cannot be robustly separated from other outflows or the diffuse emission, either spatially or kinematically. One example is the two blue-shifted lobes in region B of \ctw{}, where the lobes overlap in both projected locations and velocities (see Figures~\ref{fig:outflows_c20_b} \& \ref{fig:pv_sio}). Following the definitions in \citet{lish2020}, we classified these ambiguous identifications as `candidates'. The classifications are noted with asterisks in \autoref{tab:outflows}, Figures~\ref{fig:outflows_c20_a}--\ref{fig:outflows_sgrc_g}, and \autoref{fig:pv_sio}. Among the 43 outflows, 37 are highly likely, and 6 are candidates.

We stress that this visual identification is likely to be incomplete. Potential outflows could have been missed if they cannot be distinguished from the background emission or other outflows, or if their emission is too weak. The actual (in)completeness, however, is difficult to quantify, as the identification is based on visual inspection and is subjective in nature. Recent ALMA surveys toward infrared dark clouds in the Galactic disk using CO lines as the primary outflow tracer yield detection rates of 14\%--22\% (e.g., 62 out of 280 cores in \citealt{kong2019} and 41 out of 301 cores in \citealt{lish2020} are identified with outflows). The detection rate using SiO as the primary tracer in this paper is 4--7\% (\autoref{tab:clouds}), which is much lower. It is infeasible to directly compare, e.g., the outflow mass sensitivities of previous surveys and ours, considering that the observations use different lines as outflow tracers and assume different abundances. Assuming that the observations are sufficiently sensitive to detect all exiting outflows, the lower detection rate in our sample may suggest that we have missed a substantial number of outflows that are not traced by SiO, or may reflect the variation of outflow occurrence rates along the evolutionary stages.

Meanwhile, we also note that due to the complicated environment in the CMZ \citep[e.g., the wide-spread shock tracer emission;][]{martinpintado1997} and possible contamination from the foreground, it is likely that false positive identifications exist in our sample if such large scale shock tracer emission accidentally lies upon cores. But such accidental spatial coincidence should be rare, as the velocities of the identified outflows and cores have a continuous overlap (\autoref{fig:pv_sio}). Note that the cores associated with the outflows are all likely in the CMZ, given that the velocities of their line emission (e.g., \ceo{}; see \autoref{fig:spec} inset) are consistent with the overall velocity field in the CMZ \citep[e.g.,][]{henshaw2016}, while spiral arm clouds along the line of sight are mostly seen as absorption in \tco{} and \ceo{} (\autoref{fig:spec} inset), indicating that the overlapping spiral arm clouds consist of low-density gas and hence are unlikely to have dense cores.

SiO and SO seem to be the best outflow tracers among the six molecules. Their emission is usually well separated from the background, and is usually collimated as expected for outflows. \meth{}, \fmh{}, and HNCO often suffer from contamination from the background or foreground emission, and thus are not tracing the outflows as well as SiO and SO. \cyacet{} traces both cores and outflows. Its emission is weaker than the other molecules, therefore is not an optimized outflow tracer either. As pointed out by several previous studies, \meth{}, \fmh{}, \cyacet{}, and HNCO may be released into the gas phase by slow ($\lesssim$20~\kms{}) shocks that evaporate ice mantles of dust or be produced by gas phase reactions in post-shock regions, therefore probe the widespread low-velocity shocks in the CMZ \citep{lu2017,tanaka2018,taniguchi2018}. SiO and SO, on the other hand, may be released from the dust by sputtering of the grain core, thus better probe fast ($\gtrsim$20~\kms{}) shocks induced by the outflows.

\clearpage

\section{DISCUSSION}\label{sec:disc}

\subsection{Estimate of Physical Properties of the Outflows}\label{subsec:disc_properties}

In this section, we calculate the column densities of the six molecules detected toward the identified outflows (\autoref{subsubsec:disc_column}), introduce the method to estimate the molecular abundances in the outflows (\autoref{subsubsec:disc_abundance}), and estimate the outflow masses, and where possible, the outflow energetics (\autoref{subsubsec:disc_mass}). The uncertainties involved in the estimate of column densities, abundances, and masses are discussed in \autoref{subsubsec:disc_error}. We need an order-of-magnitude estimate on these parameters to discuss implications for astrochemistry and star formation in following sections, so the unavoidable significant uncertainties in these results (one to two orders of magnitude, as detailed in \autoref{subsubsec:disc_error}) are acceptable.

\subsubsection{Column Densities of the Shock Tracers}\label{subsubsec:disc_column}

We measure the molecular line integrated fluxes of the outflows in the primary beam corrected maps within a contour level of 3$\sigma$ for each line, and list the results in \autoref{tab:outflows}. The velocity range of the integration is chosen to start from one channel ($\sim$1.3~\kms{}) away from \vlsr{} of the core to avoid diffuse emission around the system velocity, and end at the channel where the emission drops below 2$\sigma$. The separation of 1.3~\kms{} should be able to role out most of the diffuse component as the FWHM linewidth at 0.1 pc scale in these clouds drops to this value and the linewidth at even smaller scales should be narrower \citep{kauffmann2017a}. The integrated fluxes should be lower limits given limited sensitivities and potential missed flux by the interferometer.

We then derive column densities with the measured fluxes, using the calcu toolkit\footnote{\url{https://github.com/ShanghuoLi/calcu}} \citep{lish2020}. Local thermodynamic equilibrium (LTE) conditions and optically thin line emission has been assumed. A constant excitation temperature of 70~K, which is the characteristic gas kinetic temperature in the CMZ \citep{ao2013,ginsburg2016,krieger2017}, is assumed for all the lines. Note that the adopted temperature is different from that assumed for the dust in the cores in \citetalias{lu2020a}, 20~K. Observations show that the gas temperature at $\sim$0.1~pc scale in CMZ clouds is 70~K or even higher \citep{mills2013,lu2017}, while the dust temperature at this scale is largely unconstrained \citepalias[see discussion in][]{lu2020a}. Nevertheless, in \autoref{subsubsec:disc_error} we find that the derived column densities are not sensitive to the choice of temperatures. Details of the column density calculation can be found in \autoref{appd_sec:column}.

We also select reference positions in the outflow lobes, and calculate the column densities. The reference positions must have detectable \cyacet{} emission, which we choose as the anchor molecule in the next section. The positions are adjusted to include emission of as many shock tracers as possible. A circle of 0\farcs{3} across, comparable to the synthesized beam size, is used to define the area. The reference positions are marked by black crosses in Figures~\ref{fig:outflows_c20_a}--\ref{fig:outflows_sgrc_g}, and the derived column densities are listed in \autoref{tab:outflows}. The results are used to infer molecular abundances in the next section, and are further discussed in \autoref{subsec:disc_chem} in terms of astrochemical implications.

\subsubsection{Molecular Abundances}\label{subsubsec:disc_abundance}

In order to obtain the masses of the outflows, we must adopt an abundance for each shock tracer. However, the abundances of these molecules are known to be highly variable, especially toward the environment of our outflow sample with both strong shocks and complicated factors in the CMZ. One example is SiO, whose abundance has been found to vary by a factor of $>$$10^5$ across different regions \citep[e.g.,][]{martinpintado1997,sanhueza2012,csengeri2016,li2019}.

The dust emission is not always detected toward the outflows, so anchor molecules with relatively well-constrained abundances are often used to determine abundances of other molecules. One commonly used anchor molecule is CO and its isotopologues \citep[e.g.,][]{feng2016b}. By assuming a canonical $^{12}$CO abundance of 10$^{-4}$ with respect to H$_2$ \citep{blake1987}, one might in principle use CO emission to derive the H$_2$ column density at a reference position in the outflow, and then determine the abundances of other molecules by dividing their column densities by the H$_2$ column density. However, in our data, the CO lines suffer from strong absorption and missing flux, and more importantly, morphologically they are not tracing the outflows seen in other molecules (Figures~\ref{fig:c20kms}--\ref{fig:sgrc}), probably because they are optically thick thus tracing the surface of the clouds instead of the outflows in the interior.

Therefore, we choose to anchor our estimate of molecular abundances on \cyacet{}, which is detected toward both cores and outflows and has shown a relatively stable abundance between the core/outflow environments in other observations \citep[e.g., a factor of $\sim$30 enhancement from cores to outflow lobes around low-mass and high-mass protostars;][]{bachiller1997,feng2016b,mendoza2018}. The other molecules either show up only toward the outflows (e.g., SiO, SO) or suffer from contamination by pc-scale diffuse emission or strong absorption toward the cores (e.g., \meth{}, \fmh{}, and HNCO), and thus are not appropriate as the anchor tracer.

First, we compare the column densities of \cyacet{} and of H$_2$ toward the cores, and estimate abundances of \cyacet{} in the cores. The column densities of \cyacet{} is derived following the procedures in \autoref{subsubsec:disc_column}. The H$_2$ column densities are derived using the dust continuum from \citetalias{lu2020a}. Then we assume an enhancement of factor 30 \citep{bachiller1997,feng2016b}, and obtain the abundances of \cyacet{} in the outflows. Finally we derive the H$_2$ column density in the outflows and use it to calibrate the abundances of the other shock tracers. The adopted molecular abundances with respect to molecular hydrogen are listed in \autoref{tab:outflows}, and the mean values of individual clouds are given in \autoref{tab:clouds}.

The estimated \cyacet{} abundances in the outflows are consistent in terms of the order of magnitude with previous result toward \ctw{} using multiple \cyacet{} transitions \citep[$10^{-9}$--$10^{-8}$ depending on the assumptions;][]{walmsley1986}.

We note that the estimated abundances of the molecules span a large range. For example, the SiO abundance with respect to H$_2$ in our outflow sample ranges from 10$^{-10}$ to 10$^{-8}$, with a mean value of 2.05$\times$10$^{-9}$. This justifies our choice of estimating molecular abundances case by case, rather than assuming a constant abundance, for the latter case will bias the mass estimates significantly.

There are several cases where we cannot directly determine the abundance of a molecule in an outflow, and then we have to circumvent them by making reasonable assumptions: i) One lobe of an outflow has a well-defined reference position and the abundance can be determined from a scaling of the \cyacet{} emission, while the other lobe does not. In this case, we assume that the abundances of the blue/red-shifted lobes of the same outflow are identical and adopt the abundance of the other lobe. ii) An outflow has well-defined reference positions in multiple molecular line emission including \cyacet{}, but the core associated with it does not have detectable \cyacet{} emission, and therefore the abundance of \cyacet{} in the outflow cannot be determined. In this case, we adopt a mean \cyacet{} abundance of all the other outflows in the region, and then use it to determine abundances of other molecules in this outflow. iii) An outflow has well-defined reference positions in multiple molecular line emission, but not in \cyacet{}. We have to adopt mean molecular abundances of all the other outflows in the region for each of the detected molecules in the outflow. All these cases are explicitly noted in \autoref{tab:outflows}.

\begin{figure}[!t]
\centering
\includegraphics[width=0.48\textwidth]{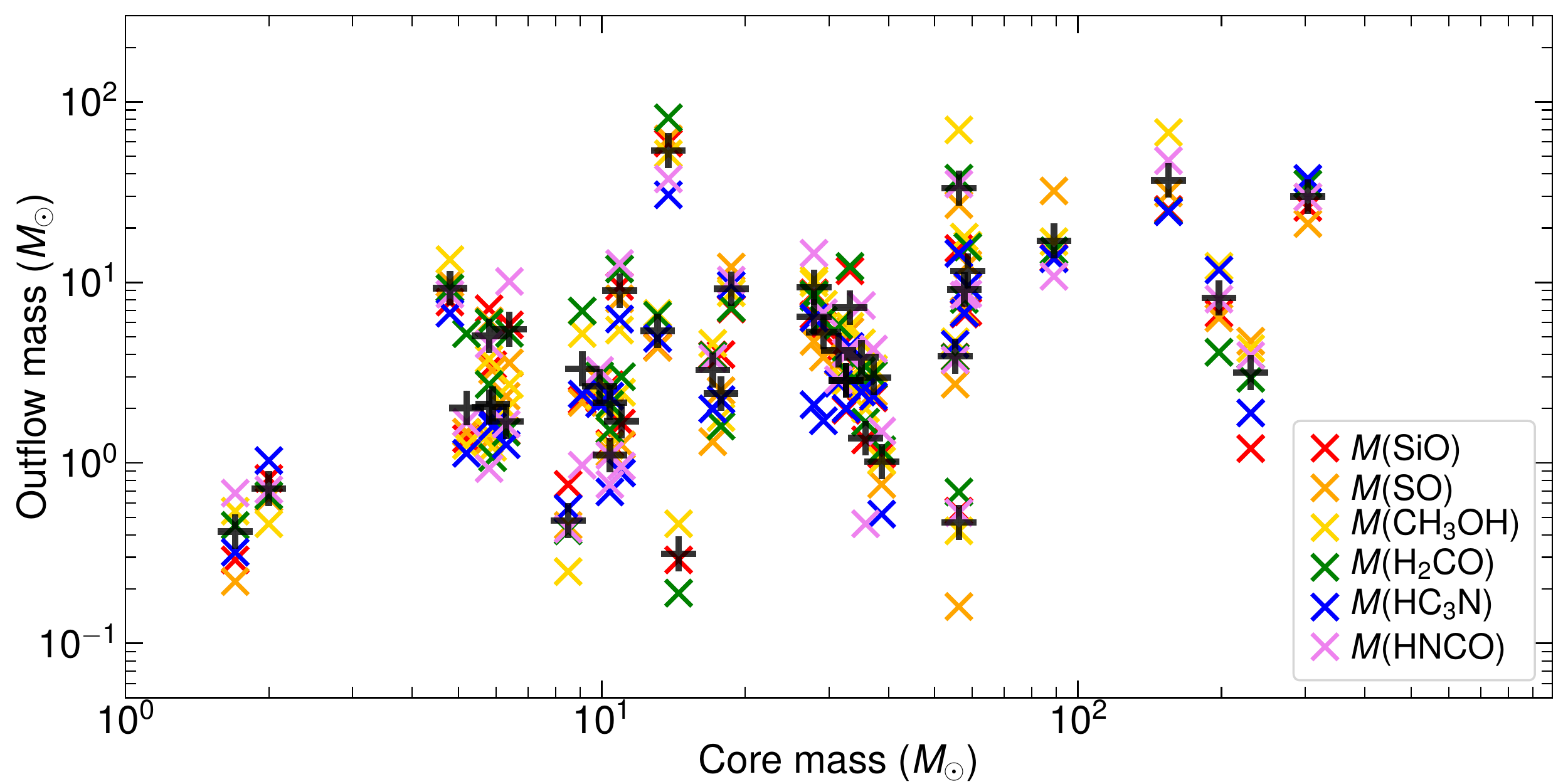}
\caption{Outflow masses derived from different molecules, plotted against masses of the cores where the outflows originate. The outflow masses are the sum of those of the blue and red lobes, color-coded by molecular tracers based on which the masses are estimated. The systematic uncertainty of factor 70 in the outflow masses is not plotted here. The black crosses denote the mean outflow masses of the different molecules.}
\label{fig:moutflow}
\end{figure}

\subsubsection{Masses and Energetics of the Outflows}\label{subsubsec:disc_mass}

After the molecular column densities and abundances are derived, the outflow masses are estimated by assuming uniform abundances within each outflow, using the calcu toolkit. The results are listed in \autoref{tab:outflows}, and plotted in \autoref{fig:moutflow}. For the same outflow lobe, multiple shock tracers could be detected, in which case we derive more than one outflow masses. All of these masses are deemed to be worth reporting, as the different molecules may trace different components of the same outflow with different chemical environments or excitation conditions. Nevertheless, the masses of the same outflow are consistent within an order of magnitude as demonstrated in \autoref{fig:moutflow}.

In a few cases where the outflow emission can be unambiguously separated from contaminations and the lobes are well collimated, we are able to measure the projected length and estimate a dynamical time-scale. Then the energetics of the outflows, e.g., the outflow rate and the outflow mechanical force, can be estimated following procedures in \citet{lish2020}. One example of such well-defined outflows is the blue-shifted lobe of the Sgr~C region F \#2 outflow. With a projected angular scale of 20\arcsec{} ($\sim$0.8~pc), and an outflow terminal velocity (the maximum velocity difference between the outflow and the core, as noted by the velocity range listed in \autoref{tab:outflows}) of 33.7~\kms{}, the dynamical time-scale is $3.2$$\times$$10^4$/$\cos$($\theta$)~yr where $\theta$ is the inclination angle of the outflow lobe with respect to the plane of the sky. The outflow mass rate is then $\sim$$6$$\times$$10^{-4}$$\cos$($\theta$)~\msolpyr{}. If the inclination angle is not too large ($\theta$$<$80\arcdeg), the outflow mass rate is $\gtrsim$$10^{-4}$~\msolpyr{}, which is usually found toward outflows around high-mass young stellar objects \citep{maud2015}. It is also possible to estimate the accretion rate, with the same assumptions in \citet{lish2020}, e.g., a wind speed of 500~\kms{} from the disk and a ratio between the accretion rate and the mass ejection rate of 3, which leads to a value of $2.5$$\times$$10^{-5}$$\cos$($\theta$)~\msolpyr{}. However, there are significant uncertainties in the outflow masses (see next section) and the true physical scale of the outflow lobes (contamination, potentially missed weak emission, inclination angle), on top of the unconstrained assumptions made in the calculation of the accretion rate \citep{lish2020}. We expect the uncertainty in the estimated outflow mass rate and accretion rate to be potentially three orders of magnitude or even greater, which is comparable to the dynamical range of our data (\autoref{fig:moutflow}). Therefore, we will not discuss these results further.

\subsubsection{Uncertainties in the Estimated Parameters}\label{subsubsec:disc_error}

There are several significant uncertainties in the estimate of column densities and outflow masses. First, we have assumed LTE conditions and a constant excitation temperature of 70~K when calculating the column densities. If we consider subthermal excitation, then the excitation temperature would be substantially lower than the kinetic temperature. If the temperature is varied between 20 and 100~K, a range that has been observed toward the CMZ \citep{ao2013,ginsburg2016,lu2017} and toward outflows in nearby clouds \citep{lefloch2012}, then the resulting mass will vary by 60\%, or 0.2~dex. Considering that the temperature could be even higher in post-shock regions \citep[$\sim$10$^3$~K;][]{tafalla2013}, this uncertainty will only be larger. Second, we have assumed optically thin emission for all the molecular lines, while this may be invalid especially for \meth{} and \fmh{}, whose optical depths are higher than the other molecular lines as evidenced by the strong self absorption toward the cores. As shown in \autoref{fig:moutflow}, the masses of the same outflow estimated from different molecules are usually consistent within an order of magnitude. Therefore, the uncertainty in the outflow masses stemming from different optical depths of the molecular lines is estimated to be half of this range at most, or 0.5~dex. Third, as discussed in \autoref{subsec:obs_combine}, the missing flux issue of the ALMA data may lead to an underestimate of 40\% for the measured flux, or 0.15~dex. The three uncertainties above go into the calculation of column densities.

For outflow masses, the molecular abundances must be taken into account additionally. The abundances of these shock tracers hinge on that of \cyacet{}, which is again based on: i) the total molecular column densities in the cores that are derived from the dust continuum, with dependences on the assumed dust temperature, the dust opacity, and the gas-to-dust mass ratio \citepalias[for detailed discussion, see][]{lu2020a}, ii) the LTE condition assumed for the calculation of \cyacet{} column densities in the cores, which may not hold given the non-LTE excitation found in several CMZ clouds \citep{mills2018}, and iii) the assumed enhancement of factor 30 of \cyacet{} from the core to the outflows. The estimated molecular abundances in the outflows usually span two orders of magnitude (\autoref{tab:outflows}). Therefore, we assign an uncertainty of one order of magnitude to the abundances, and note that the uncertainty is likely even greater.

Taken together, we estimate an uncertainty of 0.85~dex (a factor of 7) in the column densities, and an uncertainty of 1.85~dex (a factor of 70) in the outflow masses. These estimates are typical uncertainties for the individual outflows, though for particular outflows the uncertainties may be smaller or larger (e.g., for outflows where molecular abundances cannot be determined so mean abundances of the region are adopted, the uncertainties in the abundance and therefore the masses would be larger; for spatially compact outflows, the underestimate of the fluxes because of the missing flux issue would be less significant; for outflows with temperatures higher than 100~K, the uncertainties in the column densities and masses would be larger).


\begin{figure}[!t]
\centering
\includegraphics[width=0.48\textwidth]{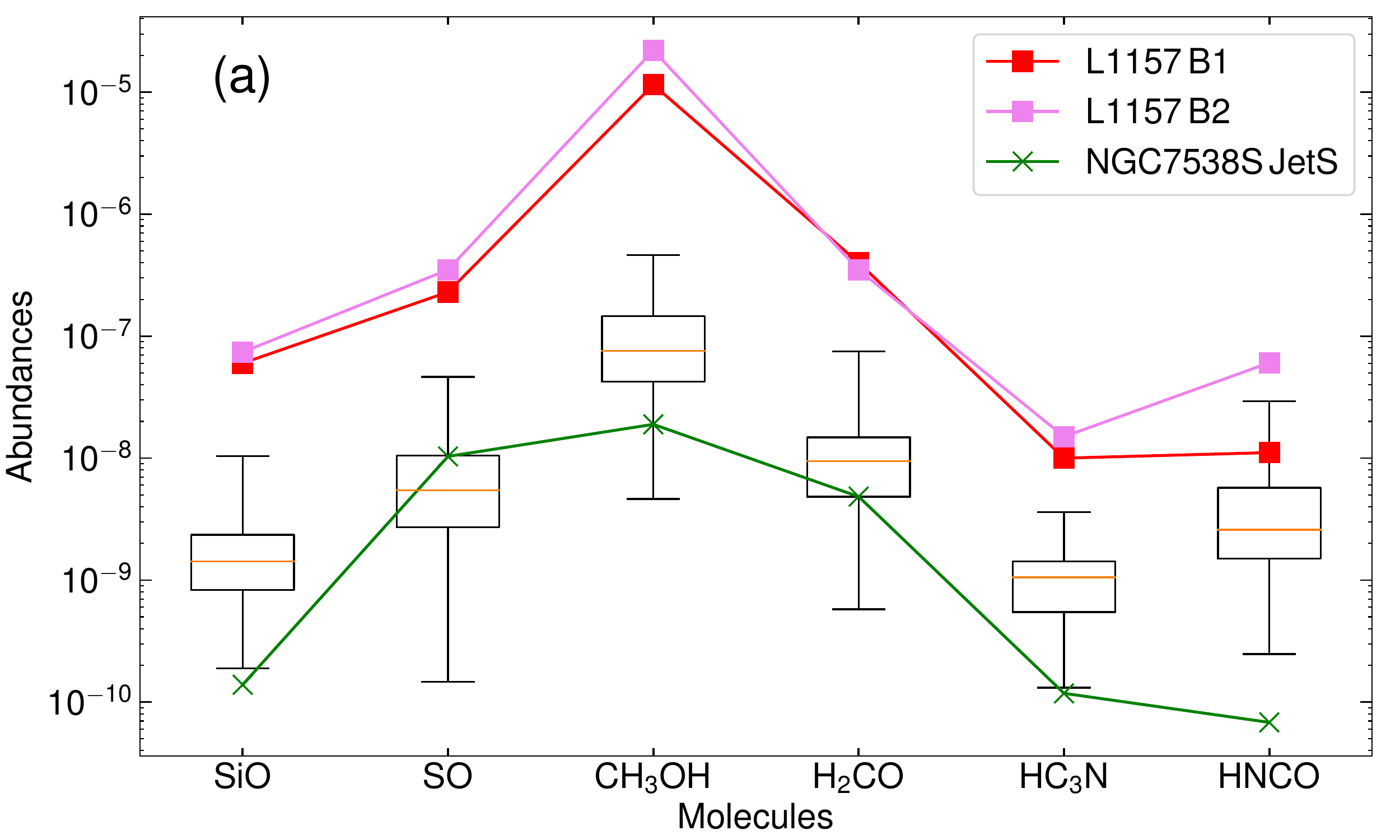} \\
\includegraphics[width=0.48\textwidth]{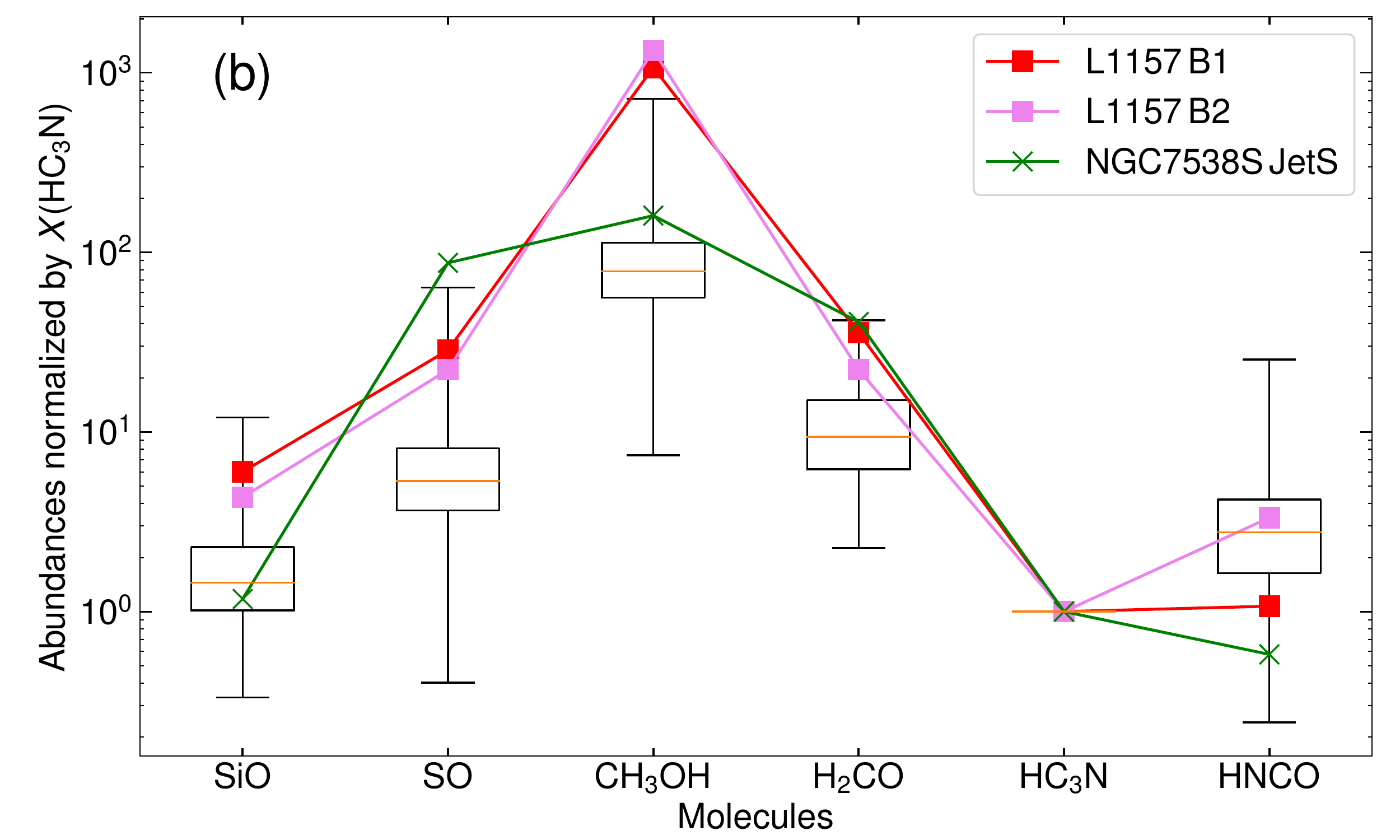}
\caption{Absolute abundances of molecules with respect to H$_2$ in the outflows are plotted in (a), and those normalized with respect to the abundance of \cyacet{} are plotted in (b). Here we only consider the abundances with independent measurements but exclude those guessed from other outflows  (i.e., entries with notes in the last column in \autoref{tab:outflows}). The boxes denote the first to third quartiles while the caps mark the full range of abundances in our outflow sample. The median of abundances of each molecule is marked by a horizontal orange line. The abundances of the low-mass outflow in L1157 and the high-mass outflow in NGC7538S are also plotted. The systematic uncertainties in the abundances are not plotted here.}
\label{fig:chem}
\end{figure}

\subsection{Shock Chemistry in the Outflows}\label{subsec:disc_chem}

We compare the relative abundances of the six shock tracers and investigate the shock chemistry in the outflows. This is the first spatially resolved astrochemical study toward outflows in the CMZ, and one of the very few such studies even including works toward Galactic disk targets.

The abundances of the six shock tracers are presented in \autoref{fig:chem}. For each molecule, we plot the median of its abundances in all the outflows as a horizontal orange line. 

To understand the relative abundances of the six molecules, we compare our result with similar studies toward nearby star forming clouds. However, we note that spatially resolved observations toward outflows that include all the six molecules, even for popular targets such as the Orion molecular cloud, are rare. For example, we are not able to find a paper that reports SiO, \fmh{}, or \cyacet{} column densities or abundances in the explosive outflow around Orion KL, even though results of SO, \meth{}, and HNCO are available \citep{feng2015}. In the end, we are able to find only two representative outflows in nearby clouds: L1157, a well-studied, prototypical low-mass outflow around a low-mass protostar \citep[e.g.,][]{bachiller1997,rodriguez2010,podio2017,holdship2019}, and NGC7538S, a prototypical massive outflow around a high-mass protostar \citep[e.g.,][]{naranjo2012,feng2016b}.

We take the abundances of the six molecules toward L1157-B1 and B2 (two reference positions in the blue-shifted lobe) from \citet{bachiller1997} and \citet{rodriguez2010}. For NGC7538S, we take the results from JetS, a reference position on the red-shifted side of the protostar, from \citet{feng2016b}, except for SiO that is not observed by the authors. We instead obtain the SiO abundance at the reference position using data from \citet{naranjo2012} (L.~Zapata, private communications). The JetN position in \citet{feng2016b} does not show \cyacet{} emission, and therefore only an upper limit of its abundance can be detected, which prevents an appropriate comparison to our results as we use \cyacet{} to infer the abundances of the other molecules. All the results from the above publications have assumed LTE conditions and optically thin line emission. The abundances are plotted in \autoref{fig:chem}(a).

In addition, we note that the abundances toward L1157-B1/B2 are estimated based on CO lines, which may become optically thick and therefore result in overestimated abundances for other molecules \citep{bachiller1997}. This may explain the systematically higher abundances for all the six molecules in L1157-B1/B2 than the other targets in \autoref{fig:chem}(a). To eliminate such biases, we normalize the abundances with respect to that of \cyacet{} in the outflows, and plot the relative abundances of the six molecules in \autoref{fig:chem}(b).

By comparing the two samples in \autoref{fig:chem}(b), the CMZ clouds vs.\ the two nearby clouds, we do not find clear evidence of difference between the relative abundances of the six shock tracers in the outflows. The relative abundances of the two nearby clouds usually fall within an order of magnitude apart from the medians of our CMZ outflow sample. However, given the significant uncertainty of the abundances and a limited sample from nearby clouds, it is premature to conclude any consistency between the two samples.

\subsection{Implications for Star Formation and Chemistry}\label{subsec:disc_sf}

Protostellar outflows are ubiquitously detected in star forming regions, suggesting active gas accretion around protostars \citep[e.g.,][]{shang2007,bally2016}. Here we investigate star formation and chemistry in the four massive clouds in the CMZ based on our observations of the outflows.

The first implication, obviously, is that protostellar accretion disks ubiquituously exist in these clouds, as protostellar outflows are supposedly driven by disks \citep{shang2007}. Direct observational evidence of protostellar accretion disks in the CMZ has been limited to the hot cores in the Sgr~B2 cloud \citep[e.g.,][]{hollis2003,higuchi2015}, and even for these cases the evidence is ambiguous given the complicated kinematic environments in Sgr~B2. More recently, D.\ Walker et al. (submitted, 2020) reported detection of protostellar outflows in G0.253$+$0.025 in the CMZ based on ALMA observations. Our finding of a large population of outflows \citep[except in \cfi{}, which is likely in a more evolved phase of star formation;][]{mills2011,lu2019a}, suggests that active accretion is ongoing around protostars in these CMZ clouds. 

The second implication concerns the evolutionary phases of star formation in these clouds based on the statistics of the cores with or without star formation signatures. In \citetalias{lu2020a}, we identify 834 cores at 2000 AU scales in the three CMZ clouds. Among them, only 43 are found to be associated with outflows. The remaining 791 cores are not associated with other signatures of star formation (masers, \hii{} regions) either, and therefore are candidates of starless cores (gravitationally bound and prestellar, or simply unbound). However, as mentioned in \autoref{subsec:results_outflows}, the outflow sample is very likely to be incomplete because of the subjectivity of the identification, thus potential outflows, even those with sufficiently strong emission, may have been missed. In addition, deeper observations may reveal more signatures of star formation such as weaker outflows or new masers. Therefore, the fraction of protostellar and starless cores is highly uncertain. For individual clouds, the fractions of cores associated with outflows range from 0.04 to 0.07 (\autoref{tab:clouds}), although this is unlikely to suggest any evolutionary trend among the three clouds given the small numbers of the outflow detections and the potential incompleteness of the outflow sample.

If we base our analysis on the current observations, i.e., 4--7\% of the cores identified in \citetalias{lu2020a} are protostellar (\autoref{tab:clouds}), then we may put a constraint on the evolutionary phase of the clouds. The time scale needed to enter the protostellar phase is of the order 1--2~Myr for both low-mass and high-mass star forming cores \citep{enoch2008,konyves2015,battersby2017}. This time scale is similar to the proposed lifetime of molecular clouds in the CMZ \citep{jeffreson2018,barnes2020}. Assuming that all the cores we detected will eventually evolve into the protostellar phase in a time scale of 1--2~Myr, the small fraction of the currently identified protostellar cores may suggest an age of star formation in these clouds as short as $\sim$0.05--0.1~Myr. In such case, star formation may have started only recently, if the clouds just condensed out of a more diffuse state, possibly driven by tidal compression during their arrival in the CMZ or on their orbit around the Galactic Center \citep{longmore2013b,kruijssen2015,kruijssen2019a}, or by the impact of adjacent expanding \hii{} regions \citep{kendrew2013,barnes2020}. Again, we stress that this estimate depends on the (in)completeness of the outflow sample, and the age of star formation in these clouds is likely longer as the fraction of protostellar cores is potentially higher.

The third implication is related to the result presented in \autoref{fig:moutflow}, where we find outflows from both high-mass ($>$100~\msol{}) and low-mass ($<$5~\msol{}) cores. Here the core masses have a strong dependence on the unconstrained dust temperature and may be overestimated by a factor of 3, as demonstrated in \citetalias{lu2020a}. Several of the high-mass cores are known to be forming high-mass protostars, with UC~\hii{} regions and class \textsc{ii} \meth{} masers \citep{lu2019a,lu2019b}. The low-mass cores, on the other hand, are only capable of forming low-mass stars with the current mass even assuming a high star formation efficiency of 50\%. Meanwhile, the majority of the outflow masses lie in the range of 1--10~\msol{}, albeit with a large uncertainty of factor 70. This mass range is characteristically found around high-mass protostars \citep{zhang2005,lu2018}. Some of the outflows have lower masses of $<$1~\msol{}, which are typical for low-mass star forming regions \citep{arce2010}. Therefore, considering the mass ranges of the cores and the outflows, the detected outflows likely trace a mix of high-mass and low-mass star formation. Simultaneous low-mass and high-mass star formation has been observed ubiquitously in massive clouds in the Galactic disk \citep[e.g.,][]{cyganowski2017,pillai2019,sanhueza2019}, which we now confirm to take place in the CMZ as well.

The last implication, as discussed in \autoref{subsec:disc_chem}, is about the shock chemistry in the outflows. Given the large uncertainties involved in the abundances (at least one order of magnitude), we are not able to conclude consistency between the shock chemistry in the CMZ clouds and in nearby analogs, but we do not find evidence of difference either. This is in contrast to the situation on the cloud scale of a few pc, where the chemistry in the CMZ is distinctly different from that in nearby clouds, e.g., an anomalous enhancement of complex organic molecules and shock tracers as compared to those in nearby clouds, suggesting the presence of wide-spread low-velocity shocks \citep{martinpintado1997,requenatorres2006,menten2009}. Several previous studies have pointed out that at the sub-0.1 pc scale, physical processes such as gas fragmentation and turbulent linewidths in the CMZ and in nearby regions may start to converge (e.g., \citealt{kauffmann2017a}; \citealt{lu2019a}; \citetalias{lu2020a}; D.\ Walker et al.\ submitted, 2020) despite distinct properties on larger scales. Our results set the first step toward a similar comparison of the shock chemistry in protostellar outflows in the CMZ and in nearby clouds. Multi-transition spectral line observations toward the CMZ outflows that enable a more robust estimate of column densities and abundances, and a larger sample of resolved astrochemical studies toward outflows in nearby clouds, will help clarify whether the shock chemistry in the two environments are consistent or not.

\section{CONCLUSIONS}\label{sec:conclusions}
As a follow-up of our \citetalias{lu2020a}, in which we used ALMA 1.3~mm continuum emission to study cores of 2000 AU scale in four massive clouds in the CMZ, we further use 1.3~mm molecular lines to identify protostellar outflows and investigate star formation activities associated with the cores. We choose six commonly used shock tracer molecules, including SiO, SO, \meth{}, \fmh{}, \cyacet{}, and HNCO\@. In three clouds (\ctw, \sgb{}, and Sgr~C), we identify 43 outflows traced by the six molecules, including 37 highly likely ones and 6 less likely ones that are considered as candidates. This is by far the largest sample of protostellar outflows identified in the CMZ. Then we estimate molecular abundances and masses of the outflows. Based on these findings and our previous studies (\citealt{lu2019a}; \citetalias{lu2020a}), we conclude that:
\begin{itemize}
\item We find no evidence of differences between the physics (existence of accretion disks, Jeans fragmentation) and shock chemistry (relative abundances of the six shock tracer molecules in the outflows) in the sub-0.1 pc scale in the CMZ and in nearby clouds. Although on the cloud scale of a few pc, gas in the CMZ exhibits extraordinary physical and chemical properties as compared to gas in the Galactic disk or in nearby clouds,  such as large turbulence linewidth, strong magnetic fields, and enhancements of particular molecules, in the smaller scale of $<$0.1~pc where gas starts to be self-gravitating, observed gas properties, and therefore physics and chemistry of the interstellar medium, may start to converge.
\item Based on the identified star formation signatures associated with the cores, the fraction of protostellar cores in these clouds may be as low as $\sim$5\%, which would indicate a short age of star formation of $\ll$1~Myr and a very early evolutionary phase for the three clouds, but this time scale is likely underestimated as our outflow sample is likely incomplete.
\item Some of the identified outflows have small masses of $\lesssim$1~\msol{} and are associated with low-mass cores of $\lesssim$5~\msol{}, and therefore likely trace low-mass star formation. Several high-mass outflows are associated with high-mass cores with known evidence of high-mass star formation. Therefore, low-mass and high-mass star formation are ongoing simultaneously in these clouds.
\end{itemize}

\acknowledgments
We thank the anonymous referee for helpful comments. X.L.~thanks Yuxin Lin, Luis Zapata, Luca Matr\`{a}, and Hauyu Baobab Liu for helpful discussions. X.L. thanks his family, Qinyu E and Xiaoe Lyu, for their support during the COVID-19 outbreak during which this manuscript was prepared. X.L. was supported by JSPS KAKENHI grants No.\ 18K13589 \& 20K14528. J.M.D.K.\ gratefully acknowledges funding from the Deutsche Forschungsgemeinschaft (DFG, German Research Foundation) through an Emmy Noether Research Group (grant number KR4801/1-1), the DFG Sachbeihilfe (grant number KR4801/2-1), and the SFB 881 ``The Milky Way System'' (subproject B2), as well as from the European Research Council (ERC) under the European Union's Horizon 2020 research and innovation programme via the ERC Starting Grant MUSTANG (grant agreement number 714907). C.B.\ and D.W.\ gratefully acknowledge support from the National Science Foundation under Award No.~1816715. This paper makes use of the following ALMA data: ADS/JAO.ALMA\#2016.1.00243.S\@. ALMA is a partnership of ESO (representing its member states), NSF (USA) and NINS (Japan), together with NRC (Canada), MOST and ASIAA (Taiwan), and KASI (Republic of Korea), in cooperation with the Republic of Chile. The Joint ALMA Observatory is operated by ESO, AUI/NRAO and NAOJ\@. Data analysis was in part carried out on the open-use data analysis computer system at the Astronomy Data Center (ADC) of NAOJ\@. This research has made use of NASA’s Astrophysics Data System.

\vspace{5mm}
\software{CASA \citep{mcmullin2007}, APLpy \citep{aplpy2012}, Astropy \citep{astropy2013}}, pvextractor
(\url{http://pvextractor.readthedocs.io}), calcu \citep{lish2020}

\clearpage

\appendix
\section{Calculation of Molecular Column Densities}\label{appd_sec:column}

Assuming optically thin emission, negligible background, Rayleigh-Jeans approximation, and local thermodynamic equilibrium (LTE) conditions, the beam-averaged column density of a molecule can be derived following \citet{mangum2015}:
\begin{equation}\label{appd_equ:col}
N_\text{tot}=\frac{8\pi k_\text{B}\nu^2}{hc^3A_{ul}}\frac{Q_\text{rot}}{g_Jg_Kg_I}\exp\left(\frac{E_u}{k_\text{B}T_\text{ex}}\right)\int T_B\text{d}v,
\end{equation}
where $k_\text{B}$ is the Boltzmann constant, $\nu$ is the rest frequency of the transition,$h$ is the Planck constant, $c$ is the speed of light, $A_{ul}$ is the spontaneous emission coefficient from the upper state $u$ to the lower state $l$, $Q_\text{rot}$ is the partition function of the molecule, $g_i$ ($i=J$, $K$, or $I$) are the degeneracies, $E_u$ is the energy of the upper state above the ground level, $T_\text{ex}$ is the excitation temperature, and $\int T_B\text{d}v$ is the integrated brightness temperature of the transition along the velocity axis. The spontaneous emission coefficient of the transition is
\begin{equation}
A_{ul}=\frac{64\pi^4\nu^3}{3hc^3}S\mu^2,
\end{equation}
where $S$ is the line strength and $\mu$ is the permanent dipole moment of the molecule. Here we directly take the values of $A_{ul}$ from the LAMDA database \citep{lamda2005}.

The partition functions $Q_\text{rot}$ are approximated using the equations in \citet{mangum2015}, as listed in \autoref{appd_tab:molinfo}. The rotation constants that are used in the calculation of $Q_\text{rot}$ are also listed in \autoref{appd_tab:molinfo}.

Then the column densities of the molecules are derived using \autoref{appd_equ:col}. The calculations are implemented in the calcu toolkit \citep{lish2020}.

\begin{deluxetable}{cccccccc}
\tabletypesize{\scriptsize}
\tablecaption{Transition spectral parameters of the outflow tracers.\label{appd_tab:molinfo}}
\tablewidth{0pt}
\tablehead{
\colhead{Molecule} & Transition & Frequency (GHz)  & $E_u/k_\text{B}$ (K) & $A_{ul}$  (s$^{-1}$) & $g_J$/$g_K$/$g_I$ & $Q_\text{rot}$ & Rotation Constants (MHz)
 }
\startdata
SiO           & 5--4                       & 217.104919 & 31.3  & 5.197$\times$$10^{-4}$  & 11/1/1  & $\frac{k_\text{B}T_\text{ex}}{hB_0}+\frac{1}{3}$ & $B_0=21711.96$ \\
\cyacet{} & 24--23                  & 218.324723 &131.0 & 0.826$\times$$10^{-3}$ & 49/1/1 & \texttt{"}  & $B_0=4549.059$ \\
SO        & 6(5)--5(4)& 219.949442 & 35.0 & 1.335$\times$$10^{-4}$ & 13/1/0.5 & \texttt{"} & $B_0=21523.556$ \\
\fmh{}     & 3(0,3)--2(0,2)    & 218.222192  & 21.0 & 2.818$\times$$10^{-4}$ & 7/1/0.25   & $\frac{1}{2}\left(\frac{\pi k_\text{B}^3T_\text{ex}^3}{h^3A_0B_0C_0}\right)^{0.5}$ & $A_0=281970.56$ \\
& & & & & & & $B_0=38833.987$  \\
& & & & & & & $C_0=34004.244$ \\
\meth{}   & 4(2) --3(1)E1      & 218.440063 & 45.5 & 4.686$\times$$10^{-5}$ & 9/1/0.25 & $\frac{1}{3}\left(\frac{\pi k_\text{B}^3T_\text{ex}^3}{h^3A_0B_0C_0}\right)^{0.5}$ & $A_0=127523.4$ \\
& & & & & & & $B_0=24690.2$  \\
& & & & & & & $C_0=23759.7$ \\
HNCO     & 10(0,10)--9(0,9) & 219.798274 & 58.0 & 1.510$\times$$10^{-4}$ & 21/1/1 & $\left(\frac{\pi k_\text{B}^3T_\text{ex}^3}{h^3A_0B_0C_0}\right)^{0.5}$ & $A_0=912711.4$ \\
& & & & & & & $B_0=11071.00$  \\
& & & & & & & $C_0=10910.57$ \\
\enddata
\end{deluxetable}

\section{Observational and Physical Properties of the Outflows}\label{appd_sec:outflowtable}
The observational and physical properties of the identified outflows are listed in \autoref{tab:outflows}.
\clearpage
\input{table_outflows.txt}

\input{OutflowsinCMZ_Jan18.bbl}
\end{CJK}
\end{document}

%% file: table_outflows.txt
\startlongtable
\begin{deluxetable*}{cccccccccc}
\tabletypesize{\scriptsize}
\tablecaption{Properties of the outflows.\label{tab:outflows}}
\tablewidth{0pt}
\tablehead{
\colhead{ID} & \vlsr{} & $M_\text{core}$ & Lobes & $\Delta v$ & $F_\text{int}$ & $N_\text{ref}$ & $X$ & $M_\text{out}$ & Notes\\
 & (\kms{}) & (\msol{}) & &  (\kms{}) & (Jy\,\kms{}) & (\sqc{}) & & (\msol{}) \\
 (1) & (2) & (3) & (4) & (5) & (6) & (7) & (8) & (9) & (10)
 }
\startdata
20\,\kms{}-A \#1 & 24.7 & 13.1& SiO-blue       & [1.5, 23.4]     & 2.83 & 1.12$\times$10$^{14}$ & 1.29$\times$10$^{-9}$ & 3.36 \\
            & [\mthc{}] &     & SO-blue        & [3.4, 23.5]     & 1.31 & 3.54$\times$10$^{14}$ & 4.08$\times$10$^{-9}$ & 3.46 \\
                 & 	    &     & \meth{}-blue   & [1.7, 23.4]     & 2.30 & 7.68$\times$10$^{15}$ & 8.84$\times$10$^{-8}$ & 4.73 \\
                 & 	    &     & \fmh{}-blue    & [2.8, 23.4]     & 1.89 & 6.32$\times$10$^{14}$ & 7.27$\times$10$^{-9}$ & 4.79 \\
                 & 	    &     & \cyacet{}-blue & [5.7, 23.4]     & 0.97 & 9.19$\times$10$^{13}$ & 1.06$\times$10$^{-9}$ & 3.96 & $X$(\cyacet) of \#2 \\
                 & 	    &     & HNCO-blue      & \nodata         & \nodata & \nodata & \nodata & \nodata \\
                 &      &     & SiO-red        & [25.8, 42.1]    & 0.79 & 5.14$\times$10$^{13}$ & 1.17$\times$10$^{-9}$ & 1.03 \\
                 & 	    &     & SO-red         & [26.0, 31.5]    & 0.38 & 2.00$\times$10$^{14}$ & 4.54$\times$10$^{-9}$ & 0.90 \\
\enddata
\tablecomments{Column (1): outflow ID. Entries marked with asterisks are candidates, and those without asterisks are highly likely outflows. Column (2): \vlsr{} of the core, and the line used to determine the \vlsr{}. Column (3): mass of the core (\citetalias{lu2020a}). Column (4): outflow lobe identifier. Column (5): velocity range of the outflow lobe. Column (6): integrated intensity of molecular emission. Column (7): column density at the reference positions, which are marked by black crosses in Figures~\ref{fig:outflows_c20_a}--\ref{fig:outflows_sgrc_g}. Column (8): adopted molecular abundance with respect to H$_\text{2}$. Column (9): outflow mass. Column (10): notes on the selection of molecular abundances.}
\end{deluxetable*}